\documentstyle{article} 
\begin{document}

\begin{center}
{\Large Stability of a relativistic rotating electron-positron jet:
nonaxisymmetric perturbations}
\end{center}

\vspace{0.5cm}
\begin{center}
{\large Ya.N.Istomin \& V.I.Pariev}
\end{center}

\begin{center}
{\it Lebedev Physical Institute, 
Leninsky Prospect 53, Moscow B-333, 117924, Russia}
\end{center}

\vspace{0.5cm}
\begin{center}
{\Large Abstract}
\end{center}

We investigate the linear stability of a hydrodynamic relativistic flow 
of magnetized plasma in the simplest case where the energy density of the 
electromagnetic fields is much greater than the energy density of the 
matter (including the rest mass energy). This is the force-free 
approximation. We considered the case of light cylindrical jet in 
cold and dense environment, so the jet boundary remains at rest.
Continuous and discrete spectra of frequencies are investigated 
analytically. An infinite sequence of eigenfrequencies is found near the 
edge of Alfv\'en continuum. Numerical calculations showed that modes 
having reasonable values of azimuthal wavenumber $m$ and radial number $n$ 
are stable and have attenuation increment $\gamma$ small. The dispersion 
curves $\omega =\omega (k_\parallel )$ have a minimum for 
$k_{{\parallel}_0}\simeq 1/R$ ($R$ is the jet radius ).  This results in 
accumulation of perturbations inside the jet with wavelength of the order 
of the jet radius. The wave crests of the perturbation pattern formed in 
such a way move along the jet with the velocity exceeding light speed.  If 
one has relativistic electrons emitting synchrotron radiation inside the 
jet, than this pattern will be visible.  This provide us with the new type 
of superluminal source. If the jet is oriented close to the line of sight, 
than the observer will see knots moving backward to the core.

\vspace{0.25cm}

{\bf Key words:} instabilities~--~MHD~--galaxies: jets~--~quasars.

\begin{center}
{\large Accepted by Mon. Not. Roy. Astron. Soc.}
\end{center}

\vspace{0.5cm}

\section{Introduction.}

Possibly the most intriguing feature of a numerous extragalactic 
radio sources is the existence of a narrow well collimated radio 
jets. It is these jets that are believed to be responsible for the 
transportation of a great amount of energy from the central compact 
areas of the galaxies to their distant radio emitting parts. From the 
observations of the superluminal motions of bright knots along the jets 
one must conclude for the velocity of the flow to be relativistic with the 
Lorentz factor $\gamma$ being of the order of 5 to 10. One of the 
important problem in the physics of extragalactic jets is their stability 
over the large distances. Up to now there have been many works in which 
the stability of the jets is investigated under the different assumptions 
for the velocity of the matter in a jet and the influence of the magnetic 
field on the flow dynamics. Turland \& Scheuer (1976) and Blandford \& 
Pringle (1976) were the first who considered the Kelvin--Helmholtz 
instability of a plane relativistic hydrodynamic flow in vortex sheet 
approximation and applied the results to extragalactic jets. The perturbed 
modes are considered both for cylindrical geometry of a jet and for the 
plane boundary between the jet and the outside media if the wavelength is 
small enough (Hardee 1979, 1987; Birkinshaw 1984; Payne \& Cohn 1985).

When it had been recognized that the magnetic field may be dynamically 
important for jet confinement, the problem of magnetohydrodynamics (MHD) 
stability of cylindrical flows naturally arose. This facet of the problem 
goes back to the investigations of the stability of magnetostatic plasma 
equilibria. Appert, Gruber \& Vaclavik (1974) have 
pointed to the existence of Alfv\'en and slow wave continua in the case of 
the sheared magnetic field and (or) nonuniform plasma density. Cohn (1983) 
found the uniform jet with the "top hat" velocity profile confined by the 
poloidal current flowing on the boundary, unstable for all modes.  He has 
also investigated relative importance for instability fundamental and 
reflection modes (these modes differ by a radial wavenumber $n$, 
fundamental mode is one having $n=0$, reflection modes have $n>0$).
The general system of the two first order differential 
equations on the radial Lagrangian displacement $\xi_{r}$ of the fluid 
element and the perturbation of the total pressure $P_{1}$, which 
describes the propagation of small disturbances along the cylindrical 
nonrelativistic MHD flow, were derived by Bondeson, Iacono \&  
Bhattacharjee (1987) (thereafter referred to as BIB).   
Interesting information concerning the existence and location of point 
eigenvalues in the complex $\omega$ plane was obtained in BIB
from a boundary layer analysis near the Suydam 
surface ${\bf kB}=0$.  Recently, Torricelli-Ciamponi \& Petrini (1990) 
and Appl \& Camenzind (1992) investigated numerically instabilities of 
nonrelativistic MHD cylindrical flows with nonzero poloidal magnetic field 
and poloidal current distributed over the jet cross section.

Next in order of complexity to relativistic hydrodynamic and 
nonrelativistic MHD approaches is relativistic MHD case, which is the 
generalization and unification both of them. When considering relativistic 
models with a significant magnetic field it is necessary to involve the 
electric force and, for non-stationary case, to add the displacement 
current. Indeed, at the speed of the 
hydrodynamic flow $v\simeq c$ the electric field in plasma with high 
conductivity is of the order of the magnetic one ${\bf E}=-{\bf v}\times 
{\bf B}/c$.  The charge density $\varrho$ is equal to $\nabla\cdot{\bf 
E}/4\pi\simeq E/L$ and the current density ${\bf j}$ in a stationary flow 
or for small nonstationarity is ${\bf j}=c\cdot(\nabla\times{\bf 
B})/4\pi\simeq cB/L$.  We see that the ratio of the electric force, 
$\varrho{\bf E}$ , acting on the unit volume, to the magnetic force, ${\bf 
j}\times{\bf B}/c$, is of the order of $E^2/B^2\simeq v^2/c^2 \simeq 1$ 
for the relativistic case. So it would be, probably, better to use the 
word electro-magnetohydrodynamics instead of magnetohydrodynamics for 
relativistic flows. Many extragalactic and, possibly, stellar jets (see 
recent discovery of superluminal motions by Mirabel \& Rodriguez, 1994) 
appear to be relativistic and magnetized, so there are numerous papers 
devoted to the construction of selfconsistent description of such flows 
(Camenzind 1987; Lovelace, Wang \& Sulkanen 1987; Li, Chiueh \& Begelman 
1992; Contopoulos 1994). It is of interest to examine the stability of the 
solutions obtained. 

We started this investigation with the simplest case where the energy 
density of the electromagnetic fields is much greater than the energy 
density of the matter (including the rest mass energy). This is 
the force-free approximation. This case is reversal to the pure 
hydrodynamics one when the electromagnetic field is absent. Using 
force-free approximation one can hope to take into account the 
influence of the electrodynamic effects on the stability and 
simultaneously obtain substantially simplified problem. 
The terms in the 
momentum equation which are proportional to the mass and the pressure of 
liquid, are therefore small compared 
to the electromagnetic force $\varrho{\bf 
E}+{\bf j}\times{\bf B}/c$, so it is possible to set $\varrho {\bf 
E}+{\bf j}\times{\bf B}/c=0$ (we use units where $c=1$). The 
force-free approximation together with the ideal hydrodynamics 
approximation (which means an infinite conductivity of plasma and 
consequently the absence of the electric field in the frame moving with 
the element of the medium) can be applied to the neighbourhood of 
the massive black hole, which is thought of as the central engine of 
active galactic nuclei. Such an approach was developed by Blandford 
\& Znajek (1977) and Macdonald (1984) [ see also chapter IV of the 
book "Black Hole: The Membrane Paradigm" by Thorne, Price \& 
Macdonald (1986) and chapter VII of the book "Physics of Black Hole" 
by Novikov \& Frolov (1986) and references therein].

In our previous paper (Istomin \& Pariev 1994, thereafter referred to as 
IP) we consider the stability of a force-free jet with respect to 
axisymmetric perturbation. Present paper is the continuation of IP and 
deals with nonaxisymmetric perturbation. As in IP the equilibrium 
configuration of the jet has cylindrical symmetry. This means that all 
quantities describing the jet depend on the distance from the jet axis $r$ 
and do not depend on the coordinate along the jet $z$ and rotational angle 
$\phi$. The boundary of the jet has the shape of a cylinder. We suggest 
that the jet propagates in the medium which density is greater than that 
of the jet but temperature and pressure are small, so the condition 
of impenetrability is fulfilled and the boundary is at rest.  The poloidal 
magnetic field $B_z$ is assumed to be uniform and parallel to the jet 
axis.  The fluid moves along spirals because of the radial electric field.
It has been proved analytically (IP) that under such conditions the 
relativistic flow is stable for axisymmetric modes ($m=0$, where $m$ is 
azimuthal wavenumber). In section~2 we derive equations governing the 
problem, describe the procedure of finding eigenfrequencies based upon the 
Laplace transformation method, and discuss asymptotic behaviour of 
perturbations in a long time after initial excitation. In section~3 
we present the results of numerical calculations.  In section~4 we perform 
boundary layer analysis of our equation near the point $r_A$ in the 
complex plane $r$, which is the point of the coincidence of two Alfv\'en 
resonant points.  Possible astrophysical implications of the results 
obtained is discussed in section~5.  

\section{Stability problem}

Let us consider a flow of liquid in a force-free cylindrical jet. 
We use the units in which $c=1$. The condition of an ideal flow is
\begin{equation}
{\bf E}=-{\bf v}\times{\bf B}\label{5},
\end{equation}
where ${\bf v}$ is the plasma velocity. The force-free approximation is
guided by the relation
\begin{equation}
\varrho{\bf E}+{\bf j}\times{\bf B}=0\label{6}.
\end{equation}

First we will review the stationary configuration of the jet described in 
IP.  In this case $\nabla\times{\bf E}=0$ and the velocity ${\bf v}$ can 
be written as 
\begin{equation} {\bf v}=K{\bf B}+\Omega^Fr{\bf e}_\phi\label{7}.  
\end{equation}
Here and below $r,z\hbox{ and }\phi$ are cylindrical coordinates, ${\bf 
e}_r$, ${\bf e}_z$ and ${\bf e}_\phi$ are the unit vectors in the 
cylindrical coordinate frame, $K=K(r)$, $\Omega^F=\Omega^F(r)$. Then 
\begin{equation}
{\bf E}=-\Omega^Fr({\bf e}_\phi\times{\bf B})\label{8}
\end{equation}
and $\Omega^F$ can be treated as
the angular rotation velocity of magnetic field lines (Thorne et al.
 1986). In the cylindrical configuration
${\bf B}=B_z(r){\bf e}_z+B_\phi(r){\bf e}_\phi$, so using Maxwell 
equations, we obtain from~(\ref{6}) 
 \begin{equation} {\bf E}(\nabla\cdot{\bf E})-{\bf 
B}\times(\nabla\times{\bf B})=0\label{9}.  
\end{equation} 
According to~(\ref{8}) the only 
non--zero component of equation~(\ref{9}) is r--component. This implies 
\begin{equation}
\Omega^FB_z\frac{d}{dr}(\Omega^Fr^2B_z)=B_z\frac{dB_z}{dr}+
\frac{1}{r}B_\phi\frac{d}{dr}(rB_\phi)\label{10}.
\end{equation}
This equation governs the force balance in radial direction and defines 
all possible solutions for the force-free
electromagnetic fields in cylindrical configuration of the magnetic tubes.
General solution of this equation was described in IP. 
For uniform poloidal magnetic field $B_z=\hbox{const}$, we have
\begin{equation}
B_\phi=\pm\,\Omega^FrB_z.\label{11}
\end{equation}
If $B_z$ is constant the stationary magnetic field 
structure is entirely determined by the function $\Omega^F(r)$.

The case where the total charge
of the jet is equal to zero is probably the most
natural. If the jet has
a charge not equal to zero, the electric field penetrates into the
surrounding medium. This results in charge motion in the plasma and in a 
decrease of the charge of the jet.  To avoid the problem of closing 
current loop somewhere outside the jet it is naturally to demand the 
total poloidal current through the jet to be equal to zero. These  two 
requirements lead to $\Omega^F(R)=0$, where $r=R$ is the jet 
boundary~(IP). 
The equilibrium stationary configuration of the jet is shown in Fig.~1, 
which is extracted from IP. We reproduce it here for illustration purpose.

\subsection{Basic equations}

We perform linear stability analysis using common method of small 
perturbations. In the subsequent formulae, the values referring to 
nonperturbed solution will be denoted by the subscript~'0', while ones 
referring to perturbation - by the subscript~'1'.
We will consider only nonaxysimmetric perturbations, so throughout the 
rest of the paper it is assumed $m\neq 0$ unless specified directly.

After removing the quantities $\varrho$, ${\bf E}$, ${\bf j}$
from the initial system of Maxwell equations and 
equations~(\ref{5})--(\ref{6}) there remains only three resultant 
equations involving ${\bf B}$ and ${\bf v}$ only:  
\begin{eqnarray}
 & & \nabla\cdot{\bf B}=0,\label{12}       \\
 & & \frac{\partial{\bf B}}{\partial t}=\nabla\times({\bf v}\times{\bf 
B}),\label{13}  \\
 & & ({\bf v}\times{\bf B}) \nabla\cdot({\bf v}\times{\bf B})-{\bf 
B}\times(\nabla\times{\bf B})- {\bf 
B}\times\nonumber\\
 & & \frac{\partial}{\partial t}({\bf 
v}\times{\bf B})=0.\label{14}  
\end{eqnarray}
The quantities ${\bf B}$ and 
${\bf v}$ can be represented as ${\bf B}={\bf B}_0+{\bf B}_1$ , 
${\bf v}={\bf v}_0+{\bf v}_1$.
We consider the perturbations in the form
\begin{eqnarray}
{\bf B}_1={\bf b}_1(r)\cdot\exp
(-i\omega t+ikz+im\phi), \nonumber\\
{\bf v}_1={\bf a}_1(r)\cdot\exp (-i\omega
t+ikz+im\phi),\label{15}
\end{eqnarray}
where $m$ is an integer.
Substitution this expressions into linearized 
set of equations~(\ref{12})--(\ref{14}) and 
removal the components of the perturbed quantities with subscript~'1' in 
favour of ${B_r}_1$ lead us to 
the following second-order ordinary differential equation of the 
variable ${B_r}_1$,
where the prime denotes differentiation with respect to $r$:
\begin{eqnarray}
 & & {B_r}_1''+{B_r}_1'\left(\frac{1}{r}-\frac{2m^2}{r^3}\frac{1}
{\omega^2-k^2-m^2/r^2}-m\frac{d\Omega^F}{dr}\frac{1}
{k+m\Omega^F}\right)+{B_r}_1\Biggl[\omega^2-k^2-\frac{m^2}{r^2}-\Biggr.\nonumber\\
 & & \left.m(k+m\Omega^F)\frac{d}{dr}\left(\frac{1}{(k+m\Omega^F)^2}
\frac{d\Omega^F}{dr}\right)\right]+
\frac{1}{k-\omega+2m\Omega^F-
{\Omega^F}^2r^2(\omega+k)}\times\label{16}  \\
 & & (A_1{B_r}_1'+A_2{B_r}_1)=0.   \nonumber
\end{eqnarray}
The quantities $A_1$ and$A_2$ are 
\begin{equation} 
A_1=-2{\Omega^F}^2r(\omega+k)+\frac{d\Omega^F}{dr}\frac{\omega+k}
{k+m\Omega^F}(m-m{\Omega^F}^2r^2-2kr^2\Omega^F) \mbox{ ,}\label{17}
\end{equation}
\begin{eqnarray}
 & & A_2=2m\left(\frac{d\Omega^F}{dr}\right)^2\frac{\Omega^Fr^2(\omega+
k)-m}{k+m\Omega^F}+\frac{d\Omega^F}{dr}\frac{1}{k+m\Omega^F} \label{18}
\left\{\frac{2m}{r}\frac{1}{\omega^2-k^2-m^2/r^2}\right.\times\nonumber\\
 & & \times\left[
\frac{m^2}{r^2}(3m\Omega^F-\omega+k)-\Omega^Fr^2(\omega^2-k^2)
\left(\Omega^F(\omega+k)+\frac{m}{r^2}\right)\right]+
\frac{m}{r}
(\omega-k)+     \\
 & & 2\Omega^Frk(\omega+k)+
\Biggl.7{\Omega^F}^2rm(\omega+k)\Biggr\}+
\frac{2m}{r^2}\Omega^F+\frac{1}{r^2}(\omega-k)+3{\Omega^F}^2(\omega+
k)   \nonumber\\
 & & +2\frac{m^2}{r^4}\frac{1}{\omega+k}
+2\frac{m^2}{r^2}\frac{1}{\omega^2-k^2-m^2/r^2}\left(
\frac{m^2}{r^4(\omega+k)}-{\Omega^F}^2(\omega+k)\right).
\nonumber
\end{eqnarray}
This equation has been obtained in IP.
When deriving this equation it was accepted that ${B_z}_0=\hbox{const}$ 
and the relation~(\ref{11}) with the sign '+' was used. The choice of the 
sign in formula~(\ref{11}) depends on the choice of the direction of the 
poloidal component of the flow in the jet.  This keeps equation~(\ref{16}) 
with definitions~(\ref{17}) and~(\ref{18}) unchanged under reversal of the 
signs of ${B_\phi}_0$ and $k$. For definiteness we adopt the sign "+" 
in~(\ref{11}) and consider arbitrary values of $k$. All other perturbed 
physical quantities can be readily expressed through ${B_r}_1$. The 
expressions for some of them are given in the Appendix~A. ${B_r}_1(r)$ 
must fulfill the boundary conditions for $r=0$ and $r=R$. For $r=0$, 
${B_r}_1$ must be regular. If $|m|\neq 1$ this condition can be 
strengthened to become ${B_r}_1\vert_{r=0}=0$.   The boundary condition 
for $r=R$ must be derived from the rigidity of the jet wall 
${v_r}_1\vert_{r=R}=0$ which has been assumed in the present 
investigation. It gives ${B_r}_1\vert_{r=R}=0$. Thus, in order to find 
perturbed modes, we have to solve the edge problem for equation~(\ref{16}) 
with the boundary conditions \begin{eqnarray} & & 
 \left.{B_r}_1\right|_{r=0} \qquad\hbox{is regular,} \nonumber\\            
 \label{19} & & \left.{B_r}_1\right|_{r=R}=0.  \end{eqnarray}

As a result of calculation parameter~$K(r)$, which determines the component 
of~${\bf v}_0$ parallel to~${\bf B}_0$, drops out from the 
equation~(\ref{16}).
Therefore, the results 
of our investigation of stability do not depend on the values and profiles 
of the longitudinal velocity. The physical reason for that is the 
neglecting in the force--balance equation~(\ref{6}), which is essentially 
the Eueler momentum equation, the terms describing the contribution from 
the inertia of the mass flow and the pressure. The results of the 
stability investigation do not depend also on the conductivity of the 
outside medium, since there are no perturbations in that region. 

The primer goal of our investigation is to answer the question whether the 
jet is stable or not. So it is more convenient to use temporal approach for
investigating stability, i.e. seek for complex values of $\omega$ for
real $k$. The existence of only one eigenvalue $\omega$ with positive 
imaginary part would mean that the jet configuration is generally unstable. 
This is not the case for spatial approach, when the search is carried out 
for complex values of $k$ for real $\omega$. In this case the sign of 
imaginary part of $k$ is irrelevant for stability and more complicated 
analysis should be done to answer the question of stability (Lifshitz \& 
Pitaevskii 1979). We adhere here to temporal approach. 

The radial eigenvalue problem~(\ref{16}) has regular singularities in the  
following 4 cases:
\begin{enumerate}
\item $k+m\Omega^F=0$ or ${\bf kB}_0=0$. This is the resonant surface 
$r=r_c$. For nonrelativistic MHD consideration  
there exist local Suydam modes in the vicinity of
$r_c$ (BIB). However this is not the
case for the force-free approximation (we see it below). Expanding 
equation~(\ref{16}) to lowest order in $x=r-r_c$, substituting ${B_r}_1 
\propto x^{\nu}$ we find the indicial equation $(\nu-1)(\nu-2)=0$. This 
implies that the solution near $r=r_c$ is the linear combination of two 
linear independent solutions, $W_1(x)$ and $W_2(x)$, with some constants 
$k_1$ and $k_2$ $\quad {B_r}_1=k_1 W_1(x)+k_2 W_2(x)$. The first terms in 
expansion of these solutions are $W_1(x)=a_1 x^2+a_2 x^3+a_3 x^4+\ldots$, 
$\quad W_2(x)=A W_1(x)\log x+b_1 x+b_2 x^2+b_3x^3+\ldots\quad$ It can be 
shown directly by inserting the expression 
for $W_2(x)$ into equation~(\ref{16}) that 
$A=0$. Therefore, this singular point does not lead to nonanalytic 
solutions in it's neighbourhood and is only apparent.
\item $\omega^2-k^2-m^2/r^2=0$. This point can be interpreted as the
resonance of the perturbation, having ${\bf k}$ parallel to magnetic
flux surfaces of ${\bf B}_0$, with the fast magnetosonic wave which in the
force-free approximation always has the speed equal to the speed of light. 
Indicial equation is $\nu(\nu-2)=0$. Similar to the type 1 singular point 
it can be shown that the coefficient of $\log x$ in the expansion of 
${B_r}_1$ near resonance vanishes, therefore this singular point is only 
apparent too. 
\item $k-\omega+2m\Omega^F-{\Omega^F}^2r^2(\omega+k)=0$. This is 
the resonance of the perturbation with the relativistic Alfv\'en wave 
$r=r_A$.  When l.h.s. has a simple zero, the indicial equation will be 
$\nu^2=0$ and, in general, ${B_r}_1$ has logarithmic singularity there. 
The expansion of other physical quantities near $r=r_A$ is given in the 
Appendix~A.
\item $r=0$. This is the singularity produced by the coordinate origin. 
Two linear independent solutions near $r=0$ are ${B_r}_1\propto r^{\mid 
m\mid-1}$ and ${B_r}_1\propto r^{-\mid m \mid-1}$, where $m\neq 0$. In 
order to meet the boundary condition~(\ref{19}) one should choose the 
former solution only.  
\end{enumerate}

One can solve edge problem for equation~(\ref{16}) directly as it was 
done in IP for the case $m=0$. However, the case $m\neq 0$ is more 
complex, particularly, because of the existence of the singularities in 
the coefficient of equation~(\ref{16}) mentioned above, two of them being 
only apparent. The numerical treatment of equation~(\ref{16}) can be 
greatly simplified and integration through 1-st and 2-nd type singular 
points can be avoided if we rewrite equation~(\ref{16}) 
as a system of the two 
first order differential equations in terms of the radial displacement 
$\xi_r$ and the disturbance of the total pressure $P_1$ instead of the 
radial component of the magnetic field perturbation ${B_r}_1$.
We need transformation into 
variables~$\xi_r$ and~$P_1$ to consider initial value problem in 
subsequent subsection and also in order to possible further generalization 
of the problem to include inertia of plasma.
For the 
Lagrangian displacement (i.e. the displacement of a fluid element moving 
with the equilibrium flow) $\mbox{\boldmath $\xi$}({\bf r},t)$ one has 
well known expression relating it to the Eulerian disturbance of the 
velocity field ${\bf v}_1$ 
\begin{equation} 
\frac{\partial \mbox{\boldmath 
$\xi$}}{\partial t}= {\bf v}_1+(\mbox{\boldmath $\xi$} \nabla){\bf v}_0- 
({\bf v}_0 \nabla)\mbox{\boldmath $\xi$}.\label{20}  
\end{equation} 
The 
$r$-component of this relation is (bearing in mind the 
representation~(\ref{15})) 
\begin{equation} 
-i\omega\xi_r={v_r}_1-ik{v_z}_0\xi_r-i\frac{m}{r}{v_\phi}_0\xi_r.\label{21}
\end{equation}
The $r$-component of the linearized induction equation~(\ref{13}) is
\begin{equation}
\left(-\omega+k{v_z}_0+\frac{m}{r}{v_\phi}_0\right){B_r}_1=\left(k{B_z}_0
+\frac{m}{r}{B_\phi}_0\right){v_r}_1.\label{22}
\end{equation}
Combining equations~(\ref{21}) and~(\ref{22}) one can 
readily obtain that ${B_r}_1=i\left(
k{B_z}_0+m/r {B_\phi}_0\right)\xi_r$, which in the force--free case with
${B_z}_0=\hbox{const}$ transforms to
\begin{equation}
{B_r}_1=i\xi_r{B_z}_0(k+m\Omega^F).\label{23}
\end{equation}
We call the value $P_0=\frac{1}{8\pi}\left({\bf B}_0^2-{\bf E}_0^2\right)$
as a total pressure because this is the diagonal part of the electromagnetic
stress tensor $\sigma_{ij}=\frac{1}{8\pi}\left(B_0^2-E_0^2\right)\delta_
{ij}-\frac{1}{4\pi}\left({B_i}_0 {B_j}_0-{E_i}_0{E_j}_0\right)$. 
Next the perturbation of the total pressure will be
\begin{equation}
P_1=\frac{1}{4\pi}\left({B_z}_0{B_z}_1+{B_\phi}_0{B_\phi}_1-{E_r}_0
{E_r}_1\right),\label{24}
\end{equation}
where the radial component of the perturbation of the electric field is
${E_r}_1=-({B_z}_0{v_\phi}_1-{B_\phi}_0{v_z}_1+{v_\phi}_0{B_z}_1-{v_z}_0
{B_\phi}_1)$ and radially directed equilibrium electric field is
${E_r}_0={v_z}_0{B_\phi}_0-{v_\phi}_0{B_z}_0=-\Omega^F r{B_z}_0$. Using 
formulas (\ref{A1})--(\ref{A3}) from the Appendix~A, 
which express ${B_\phi}_1$, ${B_z}
_1$ and ${E_r}_1$ by means of ${B_r}_1$ and $d{B_r}_1/dr$, we
find for $P_1$ expression by means of ${B_r}_1$ and $d{B_r}_1/dr$
\begin{equation}
4\pi P_1=i{B_z}_0\frac{\omega+k}{S}\left[-\frac{A}{F}\frac{d{B_r}_1}{dr}
+{B_r}_1\left\{m\frac{d\Omega^F}{dr}\frac{A}{F^2}+\frac{1}{rF}\left(
\omega-k-{\Omega^F}^2r^2(\omega+k)\right)\right\}\right],\label{25}
\end{equation}
where 
\begin{eqnarray}
 & & F=k+m\Omega^F,\label{26} \\
 & & S=\omega^2-k^2-m^2/r^2,\label{27}    \\
 & & A=k-\omega+2m\Omega^F-{\Omega^F}^2r^2(\omega+k).\label{28}  
\end{eqnarray}
Those values of $r$, which make $F$, $S$ or $A$ equal to zero, are 1-st,
2-nd or 3-d type singular points respectively. From~(\ref{25}) 
using~(\ref{23}) $d{B_r}_1/dr$ is expressed by means of $P_1$ and $\xi_r$ 
\begin{equation}
\frac{d{B_r}_1}{dr}=4\pi P_1\frac{iFS}{(\omega+k){B_z}_0A}+\frac{i}{r}
\left[\frac{F}{A}\left(\omega-k-{\Omega^F}^2r^2(\omega+k)\right)+mr
\frac{d\Omega^F}{dr}\right]{B_z}_0\xi_r.\label{29}
\end{equation}
Differentiation of the equation~(\ref{23}) and 
substitution the result for $d{B_r}_1/
dr$ in~(\ref{29}) lead after some cancellation to the first equation from 
the couple desired 
\begin{equation} 
\frac{d\xi_r}{dr}=\frac{4\pi 
P_1}{{B_z}_0^2}\frac{S}{A(\omega+k)}+\xi_r 
\frac{1}{rA}\left(\omega-k-{\Omega^F}^2r^2(\omega+k)\right).\label{30}
\end{equation}
To obtain the second equation, which relates $dP_1/dr$ to the linear 
combination of $P_1$ and $\xi_1$, we differentiate equation~(\ref{25}) on 
$r$, substitute for appeared in r.h.s. of~(\ref{25}) $d^2{B_r}_1/dr^2$ 
it's value from the second order differential equation~(\ref{16}), and 
remove $d{B_r}_1/dr$ and ${B_r}_1$ in favour of $P_1$ and $\xi_r$ by means 
of~(\ref{29}),~(\ref{23}).  Numerous cancellations occurred when this 
procedure was carrying on. Finally, we get the following equation 
\begin{equation} 
4\pi\frac{dP_1}{dr}=\frac{2\Omega^F}{A}\left(\Omega^Fr(\omega+k)-\frac{m}{r}
\right)4\pi P_1-(\omega+k)\left(A-\frac{4{\Omega^F}^2}{A}\right){B_z}_0
^2\xi_r.\label{31}
\end{equation}
We introduce for convenience the dimensionless pressure disturbance $p_\ast
=4\pi P_1/{B_z}_0^2$. Than pair of equations~(\ref{30}) and~(\ref{31}) can 
be rewritten as follows 
 \begin{eqnarray}
 & & 
 A\frac{1}{r}\frac{d}{dr}(r\xi_r)=C_1\xi_r-C_2 p_\ast,  \nonumber \\ & & 
 A\frac{dp_\ast}{dr}=C_3\xi_r-C_1 p_\ast,\label{32}
 \end{eqnarray} 
 where 
\begin{eqnarray}
C_1 & = &
\frac{2}{r}\left(m\Omega^F-{\Omega^F}^2r^2(\omega+k)\right),\label{33} \\
C_2 & = & -\frac{\omega^2-k^2-m^2/r^2}{\omega+k},\label{34}   \\
C_3 & = & -(\omega+k)(A^2-4{\Omega^F}^2).\label{35}
\end{eqnarray}

The system~(\ref{32}) has the same form as derived by Appert, Gruber \& 
Vaclavik (1974) for nonrelativistic MHD stability investigation of the 
static plasma cylinder. The only difference is in the coefficients $A$, 
$C_1$, $C_2$ and $C_3$.  The presence of the equilibrium flow does not 
influence the form of the set of the equations, but only modify the 
coefficients (e.g. BIB). Now we see that for relativistic force-free flows 
this result remains to be true.  It is easier to integrate numerically set 
of equations~(\ref{32}) than the second order equation~(\ref{16}) and in 
our numerical computations we actually integrated~(\ref{32}). First, one 
should note that no derivatives of $\Omega^F(r)$ enter into 
the~(\ref{32}).  Second, which is more important, the only singularity 
in~(\ref{32}) is $A=0$.  The values of $r$ which makes $S=0$ or $F=0$ are 
regular points of the system~(\ref{32}).  Because of the 
relation~(\ref{23}) it is clear that they will be regular points of the 
second order equation~(\ref{16}) on ${B_r}_1$ as well.  This result is 
just that we have obtained above by considering explicit expansion 
${B_r}_1$ in the neighbourhood of singular points.  

Equations~(\ref{32}) can be readily converted into one second order, so 
called Hain--L\"ust (Hain \& L\"ust, 1958) equation on $r\xi_r$ 
\begin{equation}
\frac{d}{dr}\left(\frac{A}{C_2}\frac{1}{r}\frac{d}{dr}(r\xi_r)\right)-
r\xi_r\left[\frac{d}{dr}\left(\frac{1}{r}\frac{C_1}{C_2}\right)+
\frac{C_5-C_2C_4}{rC_2}\right]=0,\label{36}
\end{equation}
where we made notice of a certain factorization
\begin{equation}
C_1^2-C_2C_3=A(C_5-C_2C_4).\label{37}
\end{equation}
The $C_4$ and $C_5$ are
\begin{eqnarray}
C_4 & = & -A(\omega+k),\label{38} \\
C_5 & = & -4{\Omega^F}^2(\omega+k).\label{39} 
\end{eqnarray}
Equations~(\ref{36}) and~(\ref{37}) are identical in form to that derived 
in BIB for nonrelativistic MHD. As well as in equation~(\ref{16}) in 
${B_r}_1$ the zero $C_2=0$ in~(\ref{36}) is only apparent singular point, 
$\xi_r$ is regular at this point. If the factorization~(\ref{37}) did not 
occur, equation~(\ref{36}) would have essential singularities when $A$ has 
a zero of quadratic or higher order.

\subsection{Laplace transformation method}

Now edge problem~(\ref{16}),~(\ref{19})  is reformulated for 
system~(\ref{32}) or equation~(\ref{36}) with evident boundary conditions: 
$\xi_r$ is finite at $r=0$ and $\xi_r=0$ at $r=R$. Thereafter we shall use 
dimensionless values $r'=r/R$, $\omega'=\omega R$, ${\Omega^F}'=\Omega^F 
R$, $k'=kR$ and the prime will be omitted. Thus, $r=1$ will correspond to 
the jet boundary.  When integrating equations~(\ref{32}) the problem 
arises how to treat the singularity $A=0$. To answer this question it is 
necessary to remember that we use temporal approach, i.e. solving initial 
value problem. We seek for the solution $\xi_r(t,{\bf r})$ for all $t>0$ 
having given the initial conditions, say, ${\bf v}_1(0,{\bf r})$ and ${\bf 
B}_1(0, {\bf r})$.  Because of the all equations governing the problem is 
linear on perturbations and the explicit dependence of this equations from 
$\phi$ and $z$ is absent, it is enough to consider the behaviour of only 
one Fourier component of all disturbances with respect to $\phi$ and $z$. 
As well as in previous text we shall take the perturbation and initial 
conditions having the form $\xi_{rkm}\propto exp(ikz+im\phi)$ and shall 
omit the subscripts $m$ and $k$ belonging to all the disturbances and 
initial conditions. Instead of doing Fourier transformation with respect 
to $t$ it is useful to apply the technique of Laplace transformation. We 
introduce function $\xi_{r\omega}(r)$ determined by
\begin{displaymath}
\xi_{r\omega}(r)=\int \limits_{0}^{\infty}\xi_r(t,r)e^{i\omega t}dt.
\end{displaymath}
Than the reverse Laplace transformation is
\begin{equation}
\xi_r(t,r)=\int \limits_{-\infty+i\sigma}^{+\infty+i\sigma}\xi_{r\omega}
(r)e^{-i\omega t}\frac{d\omega}{2\pi},\label{40}
\end{equation}
where the integration is performed along the line in the upper half of the 
complex plane $\omega$, which is parallel to the real axis and goes above 
all irregular points of $\xi_{r\omega}$. For the sake of resemblance to 
the Fourier transformation we use here the frequency $\omega$ related to 
the usual parameter $p$ of the Laplace transformation by $\omega=ip$.
The Laplace transformation of other perturbed values is defined in the 
same manner. Carrying out the Laplace transformation of the 
equation~(\ref{20}) and the source equations of the problem 
(\ref{12})--(\ref{14}), eliminating all disturbances in favour of 
$\xi_{r\omega}$, ${\bf v}_1(0,r)$ and ${\bf B}_1(0,r)$ we get finally the 
equation~(\ref{36}) with some r.h.s., which depends on the initial 
conditions.

Note, that we need to add to ${\bf v}_1(0,r)$ and ${\bf B}_1(0,r)$ the 
initial condition for the displacement $\mbox{\boldmath $\xi$} (0,r)$, 
when processing with the~(\ref{20}). Contrary to that, if we did not 
introduce displacement and did restrict ourself to Laplace analogue of 
equation~(\ref{16}) in ${B_r}_1$, the given ${\bf v}_1(0,r)$ and ${\bf 
B}_1(0,r)$ would perfectly determine the unique solution for ${\bf 
v}_1(t)$ and ${\bf B}_1(t)$.  Thus, instead of~(\ref{32}), the equations 
for $\xi_{r\omega}$ and $p_{\ast\omega}$ will be 
\begin{eqnarray}
 & & A\frac{1}{r}\frac{d}{dr}(r\xi_{r\omega})=C_1\xi_{r\omega}-C_2 
 p_{\ast\omega}+D_1,\label{41} \nonumber \\
 & & A\frac{dp_{\ast\omega}}{dr}=C_3\xi_{r\omega}-C_1 p_{\ast\omega}+D_2,
\end{eqnarray}
where $D_1$ and $D_2$ are lineary dependent on the initial conditions and 
do not contain any denominators except $\omega+k$ and Doppler shifted 
frequency $\displaystyle 
-\sigma=\omega-k{v_0}_z-\frac{m}{r}{v_0}_{\phi}= 
\omega-m\Omega^F-K{B_0}_z(k+m\Omega^F)$.  Further, we derive second order 
differential equation on $r\xi_{r\omega}$ similar to~(\ref{36}) but with 
some nonzero r.h.s.  Reduced to the normal form it will be 
\begin{eqnarray}
 & & \frac{d^2}{dr^2}(r\xi_{r\omega})+\frac{rC_2}{A}\frac{d}{dr}\left( 
 \frac{1}{r}\frac{A}{C_2}\right)\frac{d}{dr}(r\xi_{r\omega})-r\xi_{r
 \omega}\left[\frac{rC_2}{A}\frac{d}{dr}\left(\frac{1}{r}\frac{C_1}{C_2}
 \right)+\frac{C_5-C_2C_4}{A}\right]=  \nonumber\\
 & & \frac{rC_2}{A}\frac{d}{dr}\left(\frac{D_1}{C_2}\right)+\frac{r}{A^2}
 (C_1D_1-C_2D_2).\label{42}
\end{eqnarray}

To solve edge problem for~(\ref{42}) with boundary conditions 
$r\xi_{r\omega}\vert_{r=0}\propto r^{-|m|}$, $r\xi_{r\omega}\vert_{r=1}=0$
we do the following. First, integrate equation~(\ref{42}) (or, which is 
equivalently, system~(\ref{41})) from $r=0$ to $r=1$ with the initial 
condition $r\xi_{r\omega}\vert_{r=0}=r^{-|m|}$. Thus, we obtain the 
solution $g_0(r,\omega,k,m)$ of nonuniform equation~(\ref{42}), which, in 
general, does not satisfy the boundary condition at $r=1$. Then we 
integrate uniform equation~(\ref{36}) starting from $r=0$ with the same 
initial condition $r\xi_{r\omega}\vert_{r=0}=r^{-|m|}$. As a result we 
obtain some solution $g_1(r,\omega,k,m)$ of~(\ref{36}). If $g_1$ is equal 
to $0$ at $r=1$, than this will be an eigenfunction of~(\ref{36}) with the 
frequency $\omega$ being the eigenfrequency of the problem 
$\omega_{nm}(k)$. The integer number $n$ counts different eigenfrequences 
with the same $m$ and $k$.  We assume the muchness of eigenfrequences to 
be countable. This corresponds to the discret spectrum. Usually there 
exists an infinite sequence of $\omega_{nm}$ for given $m$ and $k$. This 
sequence has either finite or infinite accumulation point in 
$\omega$-plane or both (see numerical results below and section~3). The 
numbers $n$ can be chosen so that the higher $n$ the more oscillations has 
the corresponding radial eigenfunction. The least oscillating mode is 
referred to as a fundamental one in literature and the number $n=0$ is 
usually  ascribed to this mode (Birkinshaw 1984, Payne \& Cohn 1985, Appl 
\& Camenzind 1992).  All other modes are known as reflection. In the case 
$\omega=\omega_{nm}(k)$ the edge problem for the equation~(\ref{42}) has 
no solutions unless $g_0(1,\omega,k,m)=0$. If $g_0(1,\omega_{nm},k,m)=0$ 
than the number of the solutions would be infinite, because of all 
functions $g_0+cg_1$ with arbitrary constant $c$ would be equal to $0$ at 
$r=1$.  Then for $\omega$ not being the eigenfrequency the edge problem 
for~(\ref{42}) has unique solution 
\begin{equation} 
\xi_{r\omega}=g_0(r,\omega,k,m)-\frac{g_0(1,\omega,k,m)}{g_1(1,\omega,k,m)}
g_1(r,\omega,k,m).\label{43}
\end{equation}

To find out $\xi_r(t,r)$ one needs to make inverse Laplace 
transformation~(\ref{40}). The asymptotic behaviour of $\xi_r(t,r)$ for $t 
\to +\infty$ is not known a priory, therefore one can be sure that the 
definition of $\xi_{r\omega}$ is meaningful only for $\omega$ having 
$\mbox{Im}\,\omega\to +\infty$. It means that the integration 
in~(\ref{40}) must be done along the contour having $\sigma\to+\infty$ 
(long dashed line in Fig.~2).  In this region of complex $\omega$-plane 
$\xi_{r\omega}$ must be an analytical function in $\omega$ and it's values 
can be found from~(\ref{43}) by applying the procedure described above. 
Then, the contour integral~(\ref{40}) provides us with the full solution 
of the initial value problem. To answer the question of stability it is 
enough to find only the asymptotic behaviour of $\xi_r(t,r)$ for 
$t\to+\infty$. For this purpose it is necessary to continue analytically 
the expression~(\ref{43}) into the whole complex $\omega$-plane. Let us 
fixed some value of $r=r_{\ast}$ ($0<r_{\ast}<1$) and consider the 
singularities of $\xi_{r\omega}$ as a function of complex $\omega$. We 
assume that the initial conditions ${\bf v}_1(r)$, ${\bf B}_1(r)$, the 
angular rotation velocity of magnetic field lines $\Omega^F$, and, 
therefore, $D_1(r,\omega)$, $D_2(r,\omega)$ are an entire functions in the 
whole complex $r$-plane. In $\omega$-plane the only singularities of $D_1$ 
and $D_2$ are the poles $\omega=-k$ and 
$\omega=m\Omega^F+K{B_0}_z(k+m\Omega^F)$, while the coefficients in l.h.s. 
of the equation~(\ref{42}) remain to be regular at these points.

First of all, $\xi_{r\omega}$ will be irregular if $g_0(r_{\ast},\omega)$ 
or $g_1(r_{\ast},\omega)$ have singularities. It is seen from~(\ref{41}) 
that this can happens when $A(r_{\ast},\omega)=0$, i.e. when Alfv\'en 
resonant point coincides with chosen $r_{\ast}$, when $\omega=-k$, and 
when Doppler shifted frequency $\sigma$ is equal to~0 at the 
point~$r_{\ast}$. Notice, that from look at the equation~(\ref{42}) one 
might believe $C_2(r,\omega)=0$ to be the singularity of $g_0$.  However, 
it does not because of this is not the singularity of the original 
system~(\ref{41}).  From expression~(\ref{28}) for $A$ one can readily 
conclude that for any complex $\omega$ the solution $r_A(\omega)$ of 
$A(r_A,\omega)=0$ must be complex too.  Therefore, $r_A$ can be real only 
for real values of $\omega$, and $g_1$, $g_0$ are regular in the whole 
upper complex $\omega$ half--plane for any real $r$. By 
expanding~(\ref{42}) near the point $r=r_A(\omega)$ one can deduce the 
following behaviour of $g_0$ and $g_1$ in the neighbourhood of the 
Alfv\'en resonance 
\begin{eqnarray} & & g_0=c_{10}\log 
x+c_{20}+c_{30}\log^2 x+c_{40}x+c_{50}x^2+\ldots\,, \nonumber\\ & & 
 g_1=c_{11}\log x+c_{21}+c_{31}x+c_{41}x^2+\ldots\,,\label{44} 
\end{eqnarray} 
where $x=(r-r_A)/r_A$, and $c_{ij}(\omega)$ are constants 
 with respect to $r$.  We see that $r=r_A$ is the logarithmic type branch 
point of $g_1$, $g_0$ , and consequently, $\xi_{r\omega}$, considered them 
as the functions in the complex $r$-plane. On the other hand, for fixed 
$r=r_{\ast}$ and $\omega$ close to the Alfv\'en resonant frequency 
$\omega_A(r_{\ast})$ (that is $A(r_{\ast},\omega_A(r_{\ast}))=0$) the 
value $x$ can be expressed as 
\begin{equation} 
x=\frac{r'_A}{r_{\ast}}(\omega-\omega_A)+\frac{r'_A}{r_{\ast}}\left(\frac
{r^{''}_A}{2r'_A}-\frac{r'_A}{r_{\ast}}\right)(\omega-\omega_A)^2+\ldots\,,
\label{45}
\end{equation}
where
\begin{displaymath}
\left.r'_A=\frac{dr_A}{d\omega}\right|_{\omega=\omega_A(r_{\ast})}\,, \qquad
\left.r^{''}_A=\frac{d^2r_A}{d\omega^2}\right|_{\omega=\omega_A(r_{\ast})}\,.
\end{displaymath}
Inserting~(\ref{45}) into expressions~(\ref{44}) one can readily see that 
$\omega=\omega_A(r_{\ast})$ is the logarithmic type branch point for 
$g_0$, $g_1$, and, therefore, for $\xi_{r\omega}(r_{\ast},\omega)$. From 
expression~(\ref{28}) for $A$ we find
\begin{equation}
\omega_A(r_{\ast})=\frac{k+2m\Omega^F(r_{\ast})-{\Omega^F}^2(r_{\ast})
r_{\ast}^2k}{1+{\Omega^F}^2(r_{\ast})r_{\ast}^2}\,.\label{46}
\end{equation}
This equation indicates that for any real $r_{\ast}$ function 
$\xi_{r\omega}$ has unique branch point due to the Alfv\'en resonance 
$\omega_A(r_{\ast})$ and this point lies on the real axis in the complex 
$\omega$-plane. At $\omega=-k$ function $\xi_{r\omega}$ has a pole. The 
reason for this pole is the resonance of the perturbation with the 
Alfv\'en wave, propagating in the negative $z$-direction. In the case of 
uniform poloidal magnetic field ${B_z}_0=\mbox{const}$ this wave has 
always the velocity equal to the light velocity irrelevant to the value 
of~$r_{\ast}$ and to the curling angle of the magnetic field lines 
$\Omega^F r_{\ast}$ (see Appendix~B). In the case of ${B_z}_0\neq 
 \mbox{const}$ Alfv\'en waves propagating in both directions of $z$-axis 
become equal, both their velocities do depend on~$r_{\ast}$ and the second 
Alfv\'en resonance point $\omega_A(r_{\ast})$ appears. It's analytical 
properties are the same as for the first Alfv\'en resonance point: it is a 
logarithmic type branch point for  $g_0$, $g_1$, and, therefore, for 
$\xi_{r\omega}(r_{\ast},\omega)$, with the same expansion of $g_0$, $g_1$ 
near it as the expansion~(\ref{44}). It can be shown that in the general 
case of nonuniform poloidal magnetic field~${B_z}_0$ both Alfv\'en 
resonant frequencies $\omega_A(r_{\ast})$ are always real for real values 
of $r_{\ast}$ (see Appendix~B). The singularity $\sigma(r_f,\omega)=0$, or
$\omega=\omega_f(r_{\ast})=m\Omega^F+K{B_0}_z(k+m\Omega^F)$ is also real 
for any real~$r_{\ast}$. It corresponds to the convection of the 
displacement~{\boldmath$\xi$} with the fluid velocity ${\bf v}_0$.   
$\omega=\omega_f(r_{\ast})$ is a logarithmic type branch point for~$g_0$ 
and $\xi_{r\omega}(r_{\ast},\omega)$.

The second reason for irregularity of $\xi_{r\omega}$ in the complex 
$\omega$-plane arises when $r_A(\omega)=0$ of $r_A(\omega)=1$, i.e. when 
Alfv\'en resonance surface is located on the jet axis or on the jet 
boundary. In the first case it is impossible to fulfil the condition 
$r\xi_{r\omega}\mid_{r=0}\propto r^{-|m|}$ for any solution of uniform 
equation~(\ref{36}). When $r_A(\omega)=0$ and 
$\frac{dr_A}{d\omega}\vert_{r=0} \neq 0$ the behaviour of the two linear 
independent solutions near $r=0$ are 
\begin{displaymath} 
r\xi_{r\omega}\propto r^{-1/2\pm i\sqrt{4m^2-3}}.
\end{displaymath}
Therefore, in this case  any solution of~(\ref{36}) will infinitely 
oscillate and grow when approaching $r=0$. In the second case we have 
from~(\ref{44}) that in the vicinity of $r=1$ $g_0/g_1\propto \log x$, 
where $x$ is the same as in~(\ref{44}). So solution~(\ref{43}) can not 
being constructed.  The points 
\begin{equation} 
\omega_A(0)=k+2m\Omega^F(0)\label{47} 
\end{equation} 
and 
\begin{equation} 
\omega_A(1)=\frac{k+2m\Omega^F(1)-k{\Omega^F}^2(1)}{1+{\Omega^F}^2(1)} 
\label{48}
\end{equation} 
are branch type singular points of $\xi_{r\omega}$.

At last, $\xi_{r\omega}$ has also singularity whenever $\omega$ coincides 
with the eigenfrequency $\omega_{nm}(k)$. If $\omega_{nm}(k)$ is not equal 
to $\omega_A(1)$ then $g_1(1,\omega,k,m)$ can be expanded in $\omega$ into 
Taylor series in the vicinity of the point $\omega=\omega_{nm}(k)$
\begin{equation}
g_1(1,\omega,k,m)=\left.\frac{\partial g_1}{\partial \omega}\right|_{r=1}
(\omega-\omega_{nm})+\left.\frac{\partial^2g_1}{\partial \omega^2}
\right|_{r=1}(\omega-\omega_{nm})^2+\ldots\quad.\label{49}
\end{equation}
\begin{sloppypar} Now we see from general solution~(\ref{43}) that the 
function $\xi_{r\omega}$ has the pole at $\omega=\omega_{nm}(k)$ unless 
$g_0(1,\omega,k,m)=0$. If $g_0(1,\omega)$ has at $\omega=\omega_{nm}$ the 
zero in $\omega$ of the order not less than that for $g_1(1,\omega)$, then 
$r\xi_{r\omega}$ is regular at $\omega=\omega_{nm}$. The case of absence 
of the pole at eigenfrequency point takes place only for specially chosen 
initial conditions such that the eigen mode corresponding to $\omega_{nm}$ 
is not excited. For arbitrary ${\bf v}_1(0,r)$ and ${\bf B}_1(0,r)$ 
function $\xi_{r\omega}$ has poles at every eigenfrequency 
$\omega_{nm}(k)$. \end{sloppypar}

\subsection{Analytical properties of $\xi_{r\omega}$}

Let us turn to the analytical continuation of $\xi_{r\omega}(r_{\ast})$ 
from the region $\mbox{Im}\,\omega\to +\infty$ down to lower values of 
$\mbox{Im}\,\omega$. We continue to consider the Laplace transformed 
displacement $\xi_{r\omega}$ at some fixed surface $r=r_{\ast}$ inside the 
jet. From the consideration above it follows that in the upper complex 
$\omega$ half--plane $\xi_{r\omega}(r_{\ast})$ may have only poles 
$\omega_{nm}$ and singular points $r_A(\omega)$, $r_f(\omega)$ of the 
system~(\ref{41}) never become real valued. Hence, the analytical 
continuation of $\xi_{r\omega}(r_{\ast})$ into the whole upper $\omega$ 
half--plane (except poles $\omega=\omega_{nm}$) is provided by the 
formula~(\ref{43}) without any changes, i.e. when obtaining $g_1$ and 
$g_0$ integration of the system~(\ref{41}) should be performed along the 
real axis in the complex $r$-plane from $r=0$ to $r=1$. This is, however, 
not the case for further continuation of $\xi_{r\omega}(r_{\ast})$ into 
the lower $\omega$ half--plane. For $\mbox{Im}\,\omega=0$ the 
$r_A(\omega)$ cuts the real $r$ axis and moves for $\mbox{Im}\,\omega<0$ 
into the opposite complex $r$ half--plane. Note, that form 
expression~(\ref{46}) it follows that the real valued frequency always 
exists, the crossing point lying between 0 and 1.  We want to stress that 
in generally not all the points $r_A(\omega)$ cut the real $r$ axis when 
$\mbox{Im}\,\omega$ becomes 0, some of them may always lie in the upper 
$r$ half--plane, another --- in the lower half--plane. But if it occurs 
that $r_A(\omega)$ is real than the frequency $\omega$, at which it does, 
is real too. To perform the analytic continuation we must deform therefore 
the path of integration in $r$ of the system~(\ref{41}) into a contour 
extending into the complex $r$-plane and going around the point 
$r_A(\omega)$. If $r_A(\omega)$ cuts the real $r$ axis beyond the interval 
(0, 1) then integration path is not deformed. We also need to go around 
the points $r_f(\omega)$ when calculating $g_0$. After determining the 
integration path in such a way we calculate $\xi_{r\omega}(r_{\ast})$ 
according to formula~(\ref{43}), where now the functions $g_0$ and $g_1$ 
are the integrals of the~(\ref{41}) (or, equivalently, second order 
equation~(\ref{42})) along deformed path.

The procedure of analytical continuation is illustrated by Fig.~2. 
Initially integration in the formula of reverse Laplace 
transformation~(\ref{40}) is performed along long dashed line. Function 
$\xi_{r\omega}$ is continued along the paths indicated by short dashed 
lines. Letters {\sl A}, {\sl B}, {\sl C} and {\sl D} denote the paths 
corresponding to 4 possible cases of the location of their intersection 
points with real $r$ axis. Corresponding locations of Alfv\'en resonant 
points $r_A(\omega)$ and their trajectories when changing $\omega$ are 
shown on Fig.~3 by the paths  {\sl A}, {\sl B}, {\sl C} and {\sl D} 
respectively. In the cases {\sl A} and {\sl D} $r_A(\omega)$ cuts the real 
$r$ axis of the complex $r$-plane beyond the interval (0, 1) and the 
contour of integration of equations~(\ref{41}) remains to be the interval 
of the real axis (0, 1). In the case {\sl B} $r_A(\omega)$ cuts the real 
$r$ axis between $0$ and $r_{\ast}$ and drags the contour of integration. 
In the case {\sl C} cutting point lies on the interval ($r_{\ast}$, 1) and 
the contour of integration is dragged from the other side of the point 
$r_{\ast}$, at which we have to determine $\xi_{r\omega}$.

The procedure of analytical continuation described does not give unique 
results for the lower $\omega$ half--plane. The value of 
$\xi_{r\omega}(r_{\ast})$ depends on the path in $\omega$-plane along which 
the continuation is performed. Indeed, we may construct the solution 
$\xi_{r\omega}$ given by the formula~(\ref{43}) not only for real values 
$r$ but also for complex one. For this purpose when obtaining $g_0(r)$ and 
$g_1(r)$ it is necessary to perform integration along the contour in the 
complex $r$-plane, which connects three points: 0, the point $r_{\ast}$, 
at which we are interested in the $\xi_{r\omega}$, and 1. By this way we 
obtain an analytical continuation of $\xi_{r\omega}(r)$ into the complex 
$r$-plane. The only singularities of such defined $\xi_{r\omega}(r)$ are 
logarithmic type branch points $r=r_A(\omega)$ and $r=r_f(\omega)$. Hence, 
for analytical continuation of $\xi_{r\omega}(r)$ to be unique, one should 
do branch cuts attached to the points $r_A(\omega)$ and $r_f(\omega)$. 
Once chosen when $\mbox{Im}\,\omega\to 0$, these branch cuts are drawn 
with the points $r_A(\omega)$, $r_f(\omega)$ when analytical continuation 
of $\xi_{r\omega}(r)$ is performed in $\omega$. Contour of integration of 
equations~(\ref{41}) in $r$-plane must never cross them. These branch cuts 
are shown on the Fig.~3. If we find only $g_1$ (which is enough for 
determining eigenfrequencies $\omega_{nm}$), than we shall not pay 
attention to the points $r_f(\omega)$, because they are only in the r.h.s. 
of equation~(\ref{42}). It is seen that the point $r_{\ast}$ falls on 
different sides of the branch cut when $\omega$ is changed along the paths 
{\sl B} and {\sl C}.  Hence, if we continue $\xi_{r\omega}(r_{\ast})$ 
along the path {\sl C} in $\omega$-plane, we will obtain one value, while 
along the path {\sl B} the other.  Continuations along paths {\sl B} and 
{\sl A} give different results also.  The reason for this is that in the 
case {\sl B} the contour of integration goes around the singular point 
$r=r_A(\omega)$, while in the case {\sl A} it does not, therefore, we 
obtain different values of $g_0$ and $g_1$ at the point $r=1$ needed to be 
known in formula~(\ref{43}). Because of the same reason the results of the 
continuation of $\xi_{r\omega}$ along the paths {\sl D} and {\sl C} are 
also different (but they are the same for paths {\sl D} and {\sl A}).  
Thus, for the sake of $\xi_{r\omega}$ being well determined in the lower 
$\omega$ half--plane we need to cut it and keep the continuation paths 
from crossing the cuts. The cutting and choosing the continuation paths 
may be done in different ways. In our computations we choose the paths to 
be straight vertical lines in the $\omega$-plane, i.e. along the path 
$\mbox{Im}\,\omega$ runs from $+\infty$ to desired value, while 
$\mbox{Re}\,\omega$ remains fixed. The branch cuts are also straight 
vertical lines going down to infinite negative $\mbox{Im}\,\omega$. Our 
choice is reflected in Fig.~2.

Such determined $\xi_{r\omega}$ may have another singularities in the 
lower $\omega$ half--plane. First of all, there can exist the poles at 
the discrete eigenfrequences of the stability problem $\omega=\omega_{nm}$ 
with $\mbox{Im}\,\omega_{nm}\leq 0$. Secondly, the singularity may arise 
at some $\omega=\omega_{c}$ when the two points $r_A(\omega)$ merge 
together. In this case contour of integration in $r$-plane becomes 
clutched between them and the integration procedure will be undetermined. 
If we consider two paths in $\omega$-plane with slightly different 
$\mbox{Re}\,\omega$ such that $\mbox{Re}\,\omega_1<\mbox{Re}\,\omega_c<
\mbox{Re}\,\omega_2$ (paths {\sl C} and {\sl E} on Fig.~2) and plot 
corresponding trajectories of $r_A(\omega)$ in $r$-plane (see Figs.~4 (c) 
and (e) for illustration) we find that certain "reconnection" of 
trajectories occurs near the point $r_A(\omega_c)$. For 
$\mbox{Im}\,\omega<\mbox{Im}\,\omega_c$ the contour of integration catch 
on different $r_A(\omega)$ when they are moving away from merging point 
$r_A(\omega_c)$ each on it's own trajectory. Therefore, the values of 
$g_0(1)$, $g_1(1)$, $\xi_{r\omega}(r_{\ast})$ for near $\omega_1$ and
$\omega_2$ will in general strongly differ from each other, and $\omega_c$ 
will be a singular point of $\xi_{r\omega}$ with the corresponding cut to 
be attached to (see Fig.~2). We shall see below that $\omega_c$ is an 
accumulation point of the poles. It is necessary to stress that the 
singularity of $\xi_{r\omega}$ arises only if the contour of integration 
goes between the merging points $r_A(\omega)$. If it does not go around no 
one of them or bypass both than $\omega=\omega_c$ is a regular point. In 
particular, all merging points having $\mbox{Im}\,\omega_c>0$ are regular. 
The singularity may arise only for those having $\mbox{Im}\,\omega_c\leq 
0$.

\subsection{Discrete spectrum and Alfv\'en continuum}

Let us now consider the asymptotic behaviour of $\xi_r(t,r)$ at 
$t\to+\infty$. If $\xi_{r\omega}(r_{\ast})$ is defined by an analytic 
continuation into the whole cut complex $\omega$-plane the contour of 
integration in~(\ref{40}) can be deformed down to lower values of 
$\mbox{Im}\,\omega$ by going around all the poles of 
$\xi_{r\omega}(r_{\ast})$ and alongside the branch cuts (see Fig.~2). The 
part of the integral along the closing line $\mbox{Im}\,\omega\to-\infty$ 
vanishes for $t>0$ and the integral in~(\ref{40}) will be equal to the sum 
of residues of the poles at $\omega=\omega_{nm}$ and at $\omega=-k$ and 
the contributions from both sides of each cut attached to the singular 
points $\omega=\omega_b$ described above. Thus we obtain 

\begin{eqnarray}
 & & \xi_r(t,r)=i\sum_{n}\mbox{res}\vert_{\omega=\omega_{nm}}\left( 
\xi_{r\omega}e^{-i\omega t}\right)+\label{50} 
i\>\mbox{res}\vert_{\omega=-k}\left( 
\xi_{r\omega}e^{-i\omega t}\right)+ \\
 & & \frac{i}{2\pi}\sum_{\omega_b}\left[
\int\limits_{0}^{+\infty}\xi_{r\omega}^l(\omega_b-i\chi)
e^{-\chi t}d\chi-
\int\limits_{0}^{+\infty}\xi_{r\omega}^{r}(\omega_b-i\chi)e^{-\chi t}
d\chi\right]e^{-i\omega_b t},  \nonumber
\end{eqnarray}
where we have made the substitution $\omega-\omega_b=-i\chi$ and denoted 
the value of $\xi_{r\omega}$ on the left side of the branch cut as 
$\xi_{r\omega}^l$, while on the right side as $\xi_{r\omega}^r$. It is 
necessary to emphasize that the procedure of finding $\omega_{nm}$ having 
$\mbox{Im}\,\omega_{nm}<0$ depends on the arbitrary choice of the branch 
cuts, i.e. if someone do another cutting (which corresponds to another 
bypassing procedure when integrating in $r$-plane, say, to go around 
$r_A(\omega)$ in the case {\sl A} on Fig.~3), he will probably find out 
new eigenfrequences and lose the old ones. This indicates that we are not 
able to decompose the integral in~(\ref{40}) into the sum~(\ref{50}) by 
the only way:  with the cutting being changed, the contribution from poles 
may converted partially into the contribution from cuts and vice versa.  
Formula~(\ref{50}) is valid for all $t>0$. Because of the factor 
$e^{-i\omega t}$ rapidly decaying with $t$ , asymptotically for 
$t\to\infty$ the main contribution to the $\xi_r(t)$ will come from the 
poles having maximal imaginary part and from real valued $\omega_b$. Terms 
coming to expression~(\ref{50}) from the residues of the poles will be 
proportional to $\exp(-i\omega_{nm}t)$ 
\begin{equation} 
\xi_r(t,r)\vert_{\rm discr}=
\sum_{n}E_{nm}(r,k)e^{-i\omega_{nm}(k)t},\label{51}
\end{equation}
and corresponds to the discrete spectrum of $\omega$. From 
formula~(\ref{43}) and expansion~(\ref{49}) of $g_1(1,\omega)$ near the 
pole $\omega_{nm}$ it follows that the coefficients $E_{nm}(r)$ 
in~(\ref{51}) must be proportional to $g_1(r)$. Hence, $E_{nm}(r,k)$ are 
the eigenmodes for eigenfrequences $\omega_{nm}(k)$. They are the 
solutions of the edge problem for uniform equation~(\ref{36}) with the 
integration along the contour bypassing Alfv\'en resonant points 
$r=r_A(\omega)$ in complex $r$-plane as described above. Second term in 
the expression~(\ref{50}) is proportional to $\exp(ikt)$, it's radial 
profile is not an eigenmode and depends on the initial conditions. As it 
has been already mentioned, this term is the result of the degeneracy of 
one of the Alfv\'en resonant points arisen due to the 
${B_0}_z=\mbox{const}$. 

For real $r$ on the interval (0, 1), which only are meaningful for 
physics, radial eigenmodes are continuous functions when 
$\mbox{Im}\,\omega_{nm}>0$, have logarithmic type singularity at 
$r=r_A(\omega_{nm})$ when $\mbox{Im}\,\omega_{nm}=0$, and have 
a discontinuity at the point $r=r_A(\mbox{Re}\,\omega_{nm})$ of 
intersection of the branch cut with real $r$ axis when 
$\mbox{Im}\,\omega_{nm}<0$. In the last case the position of discontinuity 
depends on how we do the cutting of $r$-plane (which is related to the 
cutting of $\omega$-plane for each $0<r<1$), and is 
$r=r_A(\mbox{Re}\,\omega_{nm})$ only by our particular agreement on the 
cutting of $\omega$-plane. Radial eigenfunctions are well determined only 
for eigenfrequences with nonnegative imaginary part, i.e. for unstable or 
neutrally stable modes. This is not surprising because, as has been 
already mentioned above, one can not even uniquely determine the whole 
multitude of stable eigenfrequences $\omega_{nm}$. Such situation always 
takes place in the problem of the stability of force--free relativistic 
jet, because, as seen from~(\ref{46}), Alfv\'en resonant point lying in 
the interval (0, 1) can be always found at some real frequency $\omega$ 
for any $k$, $m\neq 0$, and for any choice of the function $\Omega^F(r)$. 
All above treatment of the initial value problem involving Laplace 
transformation shows that this conclusion on eigenmodes have to be true 
for the stability problem of any {\it ideal} hydrodynamic flow with the 
equilibrium conditions depending only on one space variable (for the 
problem being reducible to the ordinary second order differential 
equation), whenever resonant surfaces of the perturbation with the flow or 
with the characteristic waves of the medium (sonic, Alfv\'en, slow 
magnetosonic) do exist for real $\omega$. Particularly, as pointed out in 
BIB, for cylindrical nonrelativistic MHD flow singularities with the 
Alfv\'en and slow magnetosonic waves (when $A=0$ or $S=0$ in the notations 
of BIB) exist only when $\omega$ is real. Moreover, these resonances are 
the logarithmic branch points for the solutions of radial differential 
equation. Therefore, all our consideration is directly applicable to that 
case, with the only correction for more complicated structure of the 
factor ahead of $d\xi_r/dr$ and $dp_{\ast}/dr$ in BIB equations~(3) 
analogous to our equations~(\ref{32}), and, hence, more complicated 
structure of analytical continuation of $\xi_{r\omega}$ in the lower 
$\omega$ half--plane due to the enhanced number of possibilities  for 
singular points to merge together. In all previous investigations known to 
us the authors either restricted themselves to finding only unstable modes 
in their numerical computations (Torricelli-Ciamponi \& Petrini 1990, Appl 
\& Camenzind 1992) or chose some particular uniform profiles for ${\bf 
v}_0(r)$, ${\bf B}_0(r)$, equilibrium pressure $p_0(r)$ and density 
$\rho_0(r)$ such that the factor ahead of $d\xi_r/dr$ and $dp_{\ast}/dr$ 
became independent on $r$, and all resonances disappeared (Turland \& 
Scheuer 1976; Blandford \& Pringle 1976; Hardee 1979; Cohn 1983; Payne \& 
Cohn 1985). Therefore, it could be possible to integrate radial equation 
along the real $r$ axis only, and in simple cases even obtain the 
dispersion relation $D(\omega,k,m)=0$ expressed through well known 
mathematical functions. When finding stable modes it is necessary to 
bypass some of the singular points in the complex $r$-plane according to 
the procedure described above even if they are not on the real axis 
interval (0, 1).

Consider now the third term in~(\ref{50}) rising from integration along 
the branch cuts in $\omega$-plane. From~(\ref{43}), (\ref{44}) 
and~(\ref{45}) it follows that the main terms of the 
$\xi_{r\omega}(\omega)$ expansion in $\omega-\omega_A$ in the vicinity of 
the branch point $\omega_A(r)$ are 
\begin{equation} 
\xi_{r\omega}(r)=c_{30}(\omega_A)\log^2 \left[\frac{r'_A}{r}(\omega- 
\omega_A(r))\right]+\left(c_{10}(\omega_A)+\frac{g_0(1,\omega_A)}{g_1(1,
\omega_A)}\right)\log\left[\frac{r'_A}{r}(\omega-\omega_A(r))\right]+
\ldots\quad ,\label{52}
\end{equation}
with the dropped terms giving nonzero contribution to~(\ref{50}) not 
greater than $(\omega-\omega_A)\log(\omega-\omega_A)$. Substitution 
of~(\ref{51}) into~(\ref{50}) after performing calculations gives us two 
leading terms of $\xi_{r\omega}$ for $t\to\infty$ 
\begin{equation} 
\xi_r(t,r)=\left(a_1(r)\log\left(\frac{r}{r'_A(r)}t\right)+a_2(r)\right)
\frac{1}{t}e^{-i\omega_A(r)t}+\ldots\quad.\label{53}
\end{equation} 
At different points $r$ the perturbation has different frequencies 
$\omega_A(r)$ and different phase velocities 
\begin{equation}
{\bf v}_{\mbox{ph}}={\bf k}\frac{\omega_A(r)}{|{\bf k}|^2}.\label{54}
\end{equation}
This corresponds to continuous spectrum of $\omega$, since for fixed 
values $k$ and $m$ all frequencies $\omega$, with $r_A(\omega)$ lying 
between 0 and 1, are exited. As in the cases of the sheared 
noncompressible hydrodynamic flow under the presence of the critical 
surface (Timofeev 1970, for a review) and diffuse MHD pinch (Kadomtzev, 
1988) the perturbation as a whole can be treated as slitted onto  a number 
of localized, singular in $r$ perturbations propagating with the local 
Alfv\'en velocity~(\ref{54}). The decaying of perturbation observed 
from~(\ref{53}) is due to the phase mixing of neighbouring localized 
perturbation when they propagate with different velocities~(\ref{54}). 
Because for the flow considered in this paper continuous spectrum always 
present and is real the asymptotic for $t\to\infty$ behaviour of 
perturbations is determined either by an eigenfrequency with 
$\mbox{Re}\,\omega_{nm}>0$ (exponentially growing, unstable case) or by 
real eigenfrequencies if the former is absent (oscillations, neutrally 
stable case), or, at last, by continuous spectrum~(\ref{53}) (decaying 
with time, stable case) if there is no nonnegative eigenfrequencies.

\section{Results of numerical computations}

To find eigenfrequences $\omega_{nm}$ it is necessary to solve uniform 
equation~(\ref{36}). Even in the simplest case $\Omega^F=\mbox{const}$ 
this equation already has too many singularities to be solved by the well 
known hypergeometric functions and we are led to either doing numerical 
computations or finding conditions and small parameters under which it can 
be simplified. The results of numerical computations are presented in this 
section.

In accordance with the properties of $\xi_{r\omega}$ described above we 
adopted the following procedure in numerical computations. We integrated 
in $r$ pair of equation~(\ref{32}) instead of~(\ref{36}) because they have 
simpler coefficients and do not contain additional singularity at $C_2=0$.  
Integration was based upon five order Dormand--Prince method. For solving 
eigenvalue problem we applied shooting technique. For every $k$, $m$ and 
chosen initial value of $\omega$ we started from the point very close to 
$r=0$ with the initial condition $\xi_r\propto r^{|m|-1}$ and sought after 
the $\xi_r|_{r=1}$ being equal to 0. If $\mbox{Im}\,\omega>0$ integration 
should be performed entirely along the real $r$ axis. For 
$\mbox{Im}\,\omega$ negative, equal to zero or very small positive we 
passed round the zeroes of $A$ in the complex $r$-plane in accordance with 
the consideration in section~2: it is necessary to find the position of 
zeroes of $A$ in the $r$-plane for frequency $\omega_{\infty}$ having real 
part the same as $\omega$, but the imagine part very large positive, then 
plot their paths in $r$-plane when diminishing $\mbox{Im}\,\omega$ from 
$+\infty$ to the value at hand. If the path of a zero intersects with real 
$r$ axis at a point in the interval (0, 1) than this zero is needed to be 
passed round, and we have to integrate equations~(\ref{32}) along the 
contour going around the $r_A(\omega)$ (see Figs.~4 for different cases of 
paths).  In actual computations it occurred that the eigenfrequencies 
$\omega_{nm}$ have small negative imaginary part (module of it is less 
than $0.1$). If we consider only $\omega$ with $|\mbox{Im}\,\omega|\ll 1$ 
than the points $r_A(\omega)$, which are needed to be passed round, will 
lie close to the real $r$ axis. The intersection point $r_A(\omega_0)$ of 
a path of $r_A(\omega)$ with the real $r$ axis, when changing 
 $\mbox{Im}\,\omega$, will be located close to the $r_A(\omega)$. 
Therefore, numerical procedure can be simplified because there is no 
necessity to follow full path of each $r_A(\omega)$ when 
$\mbox{Im}\,\omega$ runs from $+\infty$ to the value at hand. We actually 
did the following: took real $\omega_0=\mbox{Re}\,\omega$, found all 
$r_A(\omega_0)$ belonging to the interval (0, 1), then found all 
$r_A(\omega)$ and pass round only those of them, which are the nearest to 
$r_A(\omega_0)$. In doubtful cases we did the full procedure of finding 
which points from all multitude of $r_A(\omega)$ must be passed round, 
being provided for the separate code.  The value of $\xi_r$ at $r=1$, 
which was being achieved to be 0, does not depend on the particular 
contour of integration, provided the points $r_A(\omega)$ are passed round 
properly. Hence we can suite the contour of integration for programming 
convenience. We chose the rectangular contour when going around the points 
$r_A(\omega)$, and, in order to save the accuracy of computations in the 
vicinity of singularity $r=r_A$, used to passed round also the points 
$r_A(\omega)$ when $\mbox{Im}\,\omega$ was very small positive value.

We see that for small $\mbox{Im}\,\omega_{nm}$ the rule for bypassing 
singular points in equations~(\ref{32}) coincide with that used in 
calculations of Landau damping of waves in plasma medium when doing 
integration in velocity space (Lifshitz \& Pitaevskii, 1979). There are 
two differences, however. First, in our case integration is performed in 
the interval (0, 1), while in the case of Landau damping along the whole 
real $r$ axis from $-\infty$ to $+\infty$. Second, not all singular points 
of $\xi_{r\omega}$ in $\omega$ plane are poles, but the branch points as 
well. As described in section~2 these lead to the existence of continuous 
spectrum for the problem considered in present work.     

It  occurred that $\mbox{Im}\,\omega_{nm}(k) \leq 0$ for all 
$k$, $m$, and 3 functions of $\Omega^F(r)$ involved in calculations, so  
the jet is stable with respect to helical perturbations as well 
as with respect to axisymmetric perturbations. Actually, computations 
were performed for $-10<k<10$, $-3<m<3$, and for the first 3 radial modes 
for each $k$ and $m$. The dependencies
\begin{eqnarray}
\Omega^F(r)=\Omega(1-r^2),\qquad \Omega^F(r)=\Omega e^{-r}\cos\left(\frac
{5\pi}{2}r\right),\qquad\mbox{and}  \nonumber\\
\Omega^F(r)=\Omega\left[\frac{1}{3}(r^3-1)-\frac{a+b}{2}(r^2-1)+ab(r-1)
\right]\left(\frac{a+b}{2}-ab-\frac{1}{3}\right)^{-1},    \nonumber
\end{eqnarray}
where $\Omega$, $a$, $b$ are constants, $0<a<0.5$, $0.5<b<1$, were tried 
out.  Constant $\Omega$ in the expressions for $\Omega^F$ was ranged in 
the interval from 0.1 to 20.  

Our model does not itself provide us any information about the function 
$\Omega^F(r)$ except that $\Omega^F(1)=0$. The angular rotational velocity 
of magnetic field lines has to be found from the consideration of the 
jet origin. We adhere the viewpoint that the part of the jet closest to 
the symmetry axis is connected by magnetic field lines directly to the 
black hole and can be described in the frame of force--free approximation 
(jets emerging from the accretion disks seem to be not the force--free, 
because of the kinetic energy of the mass flow is comparable to the 
Pointing flux, Pelletier \& Pudritz (1992)). Therefore, to determine 
$\Omega^F(r)$ one should solve the equations governing stationary two 
dimensional axisymmetric structure of magnetic fields in the vicinity of a 
rotating black hole taking into account the unavoidable process of 
particle creation there (Blandford \& Znajek 1977, Takahashi et al. 
1990, Nitta, Takahashi \& Tomimatsu 1991, Beskin, Istomin \& Pariev 
1992(b)).  This is still unresolved problem. Simple model describing the 
force--free magnetic field in the vicinity of a slowly rotating black hole 
was developed disregarding the effects of 
$\mbox{e}^{+}\!\mbox{e}^{-}$--pair creation in Beskin, Istomin \& Pariev, 
1992(a). The dependence 
\begin{equation} 
\Omega^F(r)=\Omega\frac{\sqrt{1-r^2}}{1+\sqrt{1-r^2}}  \label{eq:sqrt}
\end{equation}
with unspecified constant $\Omega$ follows from that model.
To be closer to reality we adopt for computations of disturbances 
propagation along the jet the function 
\begin{equation}
\Omega^F(r)=\Omega(1-r^2),  \label{eq:1-r^2}
\end{equation}
which resembles the function~(\ref{eq:sqrt}) for $0<r<1$, but is the 
entire function in complex $r$-plane, contrary to~(\ref{eq:sqrt}).

Following the procedure outlined we have calculated the dispersion 
curves $\omega_{nm}=\omega_{nm}(k)$. First three branches (n=0,1,2) of 
them for $\Omega^F$ given by~(\ref{eq:1-r^2}) and $m=2$ are shown on 
fig.~5 and fig.~6.  On fig.~7 we show the dependence of the real part of 
$\omega_{nm}(k)$ for $m=-1$.  In this case, because of the absence 
Alfv\'en resonance surface inside the jet, imaginary part of 
$\omega_{nm}(k)$ is always equal to 0 (continuum spectrum does present, 
of course).  If some 3 values $k$, $\omega$, and $m$ are a solution of 
the eigenvalue problem than the values $-k$,$-\omega^{\ast}$, and $-m$ 
will be a solution too but for the complex conjugated function 
$\xi_r^{\ast}$, so we depicted only the branches of 
$\omega_{nm}=\omega_{nm}(k)$ having $\mbox{Re}\, \omega>0$.  Those having 
$\mbox{Re}\, \omega<0$ can be obtained by the reflection of fig.~5 and 
fig.~7 with respect to the coordinates origin.  On fig.~8(a) we plotted 
an example of radial eigenmode $\xi_r$ when there is no Alfv\'en resonant 
point on the interval $0<r<1$, on fig.~8(b) the same for the case when 
resonant point exists. Note the localized character of the mode in the 
last case. In order to find physical reason for decaying of the 
discrete modes having Alfv\'en resonant point we calculated Pointing flux 
${\bf S}=1/4\pi({\bf E}\times{\bf B})$ for each mode. On fig.~8(c) the 
radial component of energy flux $rS_{r}$, calculated for mode shown on 
fig.~8(a), is depicted as a dependence from $r$. It is seen, that 
$S_{r}<0$, i.e. the energy flows to the jet axis, and, which is more 
important, Alfv\'en resonant surface is a drain of electromagnetic energy 
of the mode. This energy is converted into the energy of the mean magnetic 
and electric fields, so one should expect that near Alfv\'en resonant 
surface strong amplification of the magnetic and electric fields does  
occur. However, the consideration of this process is based on the second, 
nonlinear approximation and is beyond the scope of the present work.

\section{Boundary layer analysis}

In this section we consider the spectrum of eigenfrequencies $\omega_{nm}$ 
near the merging point $\omega_c$ of two $r_A(\omega)$. At this point 
$A(r,\omega)$ has at least quadratic zero in $r$. We assume that 
$A^{''}(\omega_c)\neq 0$ and the point $r_A(\omega_c)$ does not coincide 
with 0 and 1 (prime denotes partial differentiation with respect to $r$ at 
the point $r=r_c$). To find points $\omega_c(k,m)$ it is necessary to 
solve the couple of equations $A(r_c, \omega_c)=0$, $A'(r_c,\omega_c)=0$. 
The solution can be represented in implicit form as
\begin{eqnarray}
 & & 
k=-m\Omega^F(r_c)+\frac{m{\Omega^F}'(r_c)}{({\Omega^F}^2r^2)'|_{r=r_c}}
(1+{\Omega^F}^2(r_c)r_c^2),    \label{eq:impl1}      \\
 & & \omega_c=m\Omega^F(r_c)+ \frac{m{\Omega^F}'(r_c)}
{({\Omega^F}^2r^2)'|_{r=r_c}}(1-{\Omega^F}^2(r_c)r_c^2).  \label{eq:impl2}
\end{eqnarray}
There are two cases of the solution of the system~(\ref{eq:impl1}),
~(\ref{eq:impl2}) for each $k$ and $m$. In the first case, when the 
solution $r_c$ of (\ref{eq:impl1}) is real, $\omega_c$ determined by 
(\ref{eq:impl2}) will be real. In the second case, when (\ref{eq:impl1}) 
has a pair of complex conjugated solutions $r_c$ and $r_c^{\ast}$, 
equation~(\ref{eq:impl2}) will give us a pair of complex conjugated 
values $\omega_c$ and $\omega_c^{\ast}$. As described in Subsection~2.3, 
contour of integration of equation~(\ref{36}) can never being clamped 
between merging points $r_A(\omega)$ when $\mbox{Im}\,\omega_c>0$, so we 
are able to move it away out of the neighbourhood of $\omega_c$. Hence, 
$\omega_c$ with $\mbox{Im}\,\omega_c>0$ is regular point of 
$\xi_{r\omega}$ and can not be an accumulation point of the poles. 
Perturbations with $\mbox{Im}\, \omega<0$ are exponentially decaying with 
time and do not contribute to the asymptotic of $\xi_r(t)$ at 
$t\to\infty$. Consequently, the spectrum $\omega_{nm}$ near the real 
points $\omega_c$ is of primary importance for the stability problem. If 
$\omega_c$ and $r_c$ are real, the frequency $\omega_c$ coincides with the 
edge of the Alfv\'en continuum, i.e.  
$\left.\frac{d\omega_A(r)}{dr}\right|_{r=r_c}=0$.

Below we perform boundary layer analysis of equation~(\ref{36}) in the 
vicinity of $r_c$ similar to that was done in BIB for the edge of the slow 
wave continuum in the nonrelativistic MHD stability problem of nonrotating 
cylindrical flow. But in contrast to BIB, who always integrated radial 
eigenvalue equation along real $r$ axis, we adhere the rules of contour 
deformation in the complex $r$-plane, which are described in Section~2. At 
the point $r=r_c$, the eigenvalue problem~(\ref{36}) has a singularity 
such that to lowest order in $x=r-r_c$ 
\begin{eqnarray} 
 & & A\simeq 
 A^{''}x^2/2,  \nonumber\\ & & C_1,\  C_2, \  C_5\simeq\mbox{const}, \quad 
C_4\simeq-A^{''}(\omega+k)x^2/2,  \nonumber\\
 & & (C_1/rC_2)'\simeq\mbox{const}.  \nonumber
\end{eqnarray}
Thus, to lowest order in $x$, equation~(\ref{36}) becomes
\begin{equation}
\frac{d}{dx}\left(x^2\frac{d}{dx}(r\xi_r)\right)+D(r\xi_r)=0, 
\label{eq:outlayer}
\end{equation}
where
\begin{equation}
D=-\frac{2}{A^{''}}\left[C_5+rC_2\left(\frac{C_1}{rC_2}\right)'\right].
\label{eq:D}
\end{equation}
Setting $r\xi_r\propto x^{\nu}$, we obtain for the characteristic exponent
\begin{equation}
\nu=-1/2\pm\sqrt{1/4-D}.   \label{eq:nuexp}
\end{equation}
If
\begin{equation}
D>1/4 ,    \label{eq:osccond}
\end{equation}
than $\xi_r$ will have infinite number of oscillations in the vicinity of 
$r=r_c$ and $|\xi_r|\propto r^{-1/2}$. If $D\leq 1/4$, than two 
linear independent solutions will be of power type without oscillations. 
As we will see below, in the case $D>1/4$ there exist an infinite sequence 
of eigenfrequencies $\omega_{nm}$. Asymptotically, for $n\to\infty$, the 
eigenfrequencies converge geometrically toward the marginal point 
$\omega=\omega_c$. The eigenfunctions corresponding to these 
eigenfrequencies have the higher number of oscillations in the vicinity of 
$r=r_c$, the higher is the number $n$, and the limiting solution for 
$\xi_r$ has an infinite number of oscillations. In the other case, 
$D\leq1/4$, the picture outlined fails to be true, and the lowest--order 
boundary analysis can not provide any information on eigenfrequencies. 
Using definition~(\ref{eq:D}) and relations~(\ref{eq:impl1}),
~(\ref{eq:impl2}) oscillations condition~(\ref{eq:osccond}) can be 
transformed to the form
\begin{equation}
\frac{d}{dr}\left[\frac{{\Omega^F}'}{({\Omega^F}^2r^2)'}\right]\frac{d}{dr}
\left[r^8\frac{{\Omega^F}'}{({\Omega^F}^2r^2)'}\right]<0,
\label{eq:osc1}
\end{equation}
where all derivatives are evaluated at the point $r=r_c$. We see that this 
condition depends only on the position of the point $r_c$, and does not 
depend on $k$ and $m$ separately.

Let us now consider the frequency $\omega$ slightly different from 
$\omega_c$
\begin{equation}
\omega=\omega_c+\Delta,   \label{eq:eqn1}
\end{equation}
where $|\Delta|\ll 1$. In the vicinity of $\omega=\omega_c$, $r=r_c$ the 
main terms of expansion $A$ in $\omega$ and in $r$ are 
\begin{equation}
A=-(1+{\Omega^F}^2r^2)\Delta+\left[2m{\Omega^F}^{''}-(\omega_c+k)({\Omega^F
}^2r^2)^{''}\right]x^2/2.   \label{eq:eqn2}
\end{equation}
Note, that the equation~(\ref{eq:outlayer}) is invariant under rescaling 
of $x$. Furthermore, from (\ref{eq:eqn2}) it follows, that the rescaling 
of $x$ leaves the lowest--order in $\Delta$ and in $x$ approximation of 
radial mode equation~(\ref{36}) invariant if $\Delta$ is scaled the same 
as $x^2$. Therefore, boundary layer of thickness $x\propto\Delta^{1/2}$ 
arises near the point $r=r_c$ in the equation~(\ref{36}). To obtain "inner 
layer" equation we introduce inner rescaled variable $X$ by the following 
definition 
\begin{equation} 
X=\frac{x}{\sqrt{\Delta}}\left(-\frac{1}{D}\frac{2}{r_c}\frac{m{\Omega^F}'}
{1+{\Omega^F}^2r_c^2}\right)^{1/2},    \label{eq:eqn3}
\end{equation}
where we determine $\sqrt{\Delta}$ such that the $-\pi/2<\arg 
\sqrt{\Delta}\leq\pi/2$ and accept for the argument of the square root 
from the expression enclosed by parentheses the value 0, if this 
expression is positive, and the value $\pi/2$, if this expression is 
negative. With such definition the lower--order inner--layer equation 
becomes 
\begin{equation}
\frac{d}{dX}\left[(1-X^2)\frac{d}{dX}(r\xi_r)\right]-D(r\xi_r)=0.
\label{eq:eqn4}
\end{equation}
This is Legendre's equation with singular points $X=1$ and $X=-1$. At 
these points $r\xi_r$ has logarithmic singularities, as they are merely 
two closely spaced first--order zeroes of $A$ in $r$.  
Equation~(\ref{eq:eqn4}) shows that within the "inner layer", $|X|\ll 1$, 
the finite frequency shift is essential. Furthermore, for $|X|\gg 1$ and 
$|x|\ll 1$, there exists an "intermediate" region where the frequency 
shift becomes negligible, and where the inner--layer equation~
(\ref{eq:eqn4}) approaches the small $|x|$ limit of the external equation~
(\ref{eq:outlayer}). The boundary layer analysis consists of matching the 
solutions of the external equation~(\ref{eq:outlayer}) for $|x|$ small 
with those of the complete inner--layer equation~(\ref{eq:eqn4}) for $|X|$ 
large, using (\ref{eq:eqn3}) to relate $|x|$ and $|X|$, to obtain an 
equation for $\Delta$. Figs.~9~(a),~(b) show the complex $x$ and $X$ 
planes for the case when $\mbox{Im}\,\Delta<0$. On Fig.~9~(a) 
$m{\Omega^F}'/D<0$, on Fig.~9~(b) $m{\Omega^F}'/D>0$. For definiteness, we
assume that the contour of integration of equation~(\ref{36}) approaches 
the two merging points $r_A(\omega)$ along the real $x$ axis from $x<0$ 
and moves away them also along the real $x$ axis to $x>0$. Let us 
introduce the notation for characteristic exponent~(\ref{eq:nuexp}) 
\begin{equation} \nu=-1/2\pm is, \quad s=\sqrt{D-1/4}.  \label{eq:defs} 
\end{equation}
Parameter $s$ is real for oscillatory solutions and imagine if the 
otherwise. Then, the solution of the external equation~(\ref{eq:outlayer}) 
in the vicinity of $r=r_c$ is for $x<0$
\begin{equation}
r\xi_r=a_{+}x^{-1/2+is}+a_{-}x^{-1/2-is},   \label{eq:eqn5}
\end{equation}
where $a_{+}/a_{-}$ is fixed by the external solution in order that the 
boundary condition at $r=0$ be satisfied. Similarly, for $x>0$,
\begin{equation}
r\xi_r=b_{+}x^{-1/2+is}+b_{-}x^{-1/2-is},   \label{eq:eqn6}
\end{equation}
and $b_{+}/b_{-}$ is fixed by the external solution in order that the 
boundary condition at $r=1$ be satisfied. Matching the solution of the 
internal equation with the solution~(\ref{eq:eqn5}),~(\ref{eq:eqn6}) of 
the external equation leads to asymptotical, when $|X|\to\infty$, 
expressions for inner--layer solution
\begin{displaymath}
r\xi_r=\left\{\begin{array}{ll}
b_{+}\left(\sqrt{\tilde{\Delta}}X\right)^{-1/2+is}+
b_{-}\left(\sqrt{\tilde{\Delta}}X\right)^{-1/2-is} & \mbox{when $\mbox{Re}
\,X\to+\infty$,}  \\
a_{+}\left(\sqrt{\tilde{\Delta}}X\right)^{-1/2+is}+
a_{-}\left(\sqrt{\tilde{\Delta}}X\right)^{-1/2-is} & \mbox{when $\mbox{Re}
\,X\to-\infty$.}  \end{array}\right.
\end{displaymath}
Here we denote 
\begin{displaymath}
\tilde{\Delta}=-\Delta\frac{1}{D}\frac{2}{r_c}\frac{m{\Omega^F}'}{1+
{\Omega^F}^2r_c^2}.
\end{displaymath}
Since the scaled inner--layer equation (\ref{eq:eqn4}) is independent of 
$\Delta$, so is the connection matrix between the expansion coefficients 
for $r\xi_r$ in the two asymptotic regions $\mbox{Re}\,X\to+\infty$ and 
$\mbox{Re}\,X\to-\infty$ divided from each other by branch cuts attached 
to the points $X=+1$ and $X=-1$ (if the contour of integration goes {\it 
between} these points). Thus,
\begin{equation}
\left(\begin{array}{l}
b_{+}\left(\sqrt{\tilde{\Delta}}\right)^{is} \\
b_{-}\left(\sqrt{\tilde{\Delta}}\right)^{-is}  \end{array}\right)=
\left(\begin{array}{cc}
v_{11} & v_{12} \\ v_{21} & v_{22}
\end{array}\right)
\left(\begin{array}{l}
a_{+}\left(\sqrt{\tilde{\Delta}}\right)^{is} \\
a_{-}\left(\sqrt{\tilde{\Delta}}\right)^{-is}  \end{array}\right).
\label{eq:matr}
\end{equation}
Here $v_{ij}$ only depend on $D$. The explicit form of $v_{11}$, $v_{12}$, 
$v_{21}$, $v_{22}$ follows from the solution of the inner--layer 
equation~(\ref{eq:eqn4}). The general solution of this equation is 
\begin{displaymath}
r\xi_r=A\,P_{\nu}(X)+B\,Q_{\nu}(X),
\end{displaymath}
where $P_{\nu}(X)$ and $Q_{\nu}(X)$ are Legendre's functions of the first 
and second order respectively with $\nu$ from~(\ref{eq:defs}). The main 
terms of the asymptotic expansion of $P_{\nu}(X)$ and $Q_{\nu}(X)$, when
$|X|\to\infty$, are (Abramowitz \& Stegun, 1970)
\begin{eqnarray}
 & & P_{\nu}(X)=\frac{1}{\sqrt{\pi}}
\frac{\Gamma(-is)}{\Gamma(1/2-is)}\left(
2X\right)^{-1/2-is}+\frac{\Gamma(is)}{\Gamma(1/2+is)}\left(2X\right)^
{-1/2+is},  \nonumber\\
 & & Q_{\nu}(X)=\sqrt{\pi}\frac{\Gamma(1/2+is)}
{\Gamma(1+is)}\left(2X\right)^ 
{-1/2-is}.  \nonumber
\end{eqnarray}
The form of these expansions just match the expansion (\ref{eq:eqn5}), 
(\ref{eq:eqn6}) of the solution of the outer--layer equation. Therefore, 
considering branch cuts as shown on Figs.~9, involving the relations 
between $P_{\nu}$ and $Q_{\nu}$ on the different sides of a branch cut, 
and properly choosing the branches of functions $X^{-1/2\pm is}$ in the 
asymptotic expressions, one can obtain the coefficients of connection 
matrix (if the contour of integration goes between the points $X=+1$ and 
$X=-1$)
\begin{eqnarray}
 & & v_{11}=-e^{-2\pi s}(1+\coth\pi s), 
\quad v_{22}=e^{2\pi s}(\coth \pi s-
1), \quad v_{12}=\sqrt{\pi}se^{2\pi s}\left(\frac{\Gamma(is)}{\Gamma
(1/2+is)}\right)^2,  \nonumber\\
 & & v_{21}=-\frac{se^{-2\pi s}}{\sqrt{\pi}}
\left(\frac{\Gamma(-is)}{\Gamma(1/2
-is)}\right)^2.   \nonumber
\end{eqnarray}
Then, taking into account that $R_{+}=b_{+}/b_{-}$ and $R_{-}=a_{+}/a_{-}$
are fixed by boundary conditions, we find that in order for  the system 
(\ref{eq:matr}) to have nontrivial solution for $a_{+}$, $a_{-}$, $b_{+}$, 
$b_{-}$, the following dispersion relation must be fulfilled
\begin{displaymath}
v_{21}R_{+}R_{-}\left({\tilde{\Delta}}^{is}\right)^2+(v_{22}R_{+}-v_{11}
R_{-}){\tilde{\Delta}}^{is}-v_{12}=0.
\end{displaymath}
This is a quadratic equation on ${\tilde{\Delta}}^{is}$ with the solutions
\begin{displaymath}
{\tilde{\Delta}}^{is}_{\pm}=\zeta_{\pm}=
\frac{1}{2v_{21}\sqrt{R_{+}R_{-}}}\left[-
\frac{e^{\pi s}R_{+}+e^{-\pi s}R_{-}}{\sqrt{R_{+}R_{-}}\sinh\pi s}\pm
\sqrt{\left(\frac{e^{\pi s}R_{+}+e^{-\pi s}R_{-}}
{\sqrt{R_{+}R_{-}}\sinh\pi s}\right)^2-4\frac{\cosh^2\pi s}{\sinh^2\pi s}}
\right].
\end{displaymath}
Because of the multivaluedness of the raising to a complex power there 
exist two infinite sequences of the solutions $\tilde{\Delta}$
\begin{equation}
{\tilde{\Delta}}_{n(\pm)}=\exp\left[\frac{1}{s}\arg\zeta_{\pm}-\frac{2\pi 
n}{s}-\frac{i}{s}\log|\zeta_{\pm}|\right],   \label{eq:seq}
\end{equation}
where $n$ is an integer number. If $s$ is real ($s>0$) than with 
increasing $n$ the module of $\Delta_{n(\pm)}$ will be decreasing and the 
sequences of eigenfrequencies $\omega_{nm}=\omega_c+\Delta_{n(\pm)}$ will 
converge to marginal point $\omega_c$. Note, that the analysis in this 
section is only valid for small $\Delta$, i.e. the 
expression~(\ref{eq:seq}) is accurate only when $n$ is large and positive. 
If $s$ is imagine and oscillations condition~(\ref{eq:osccond}) breaks 
down, this will not be the case. If $D<1/4$ than $\Delta_{n(+)}$ and 
$\Delta_{n(-)}$ from~(\ref{eq:seq}) have the same modules for every $n$, 
with only the argument changing with $n$. Hence, these eigenfrequencies 
are improper, and in reality there are no eigenfrequencies in the vicinity 
of $\omega_c$.

Without knowledge about $R_{+}$ and $R_{-}$ one could say nothing about 
the sequences of $\Delta_{n(\pm)}$ except their existence. In turn, the 
values of $R_{+}$ and $R_{-}$ depend on the solution of full 
equation~(\ref{36}) with possibly existing another points $r_A(\omega)$ 
needed to be gone round when integrating~(\ref{36}).

At last, we mention that in the case when contour of integration does not 
go between the points $X=+1$ and $X=-1$, but go round the whole couple, 
than the coefficients of the connection matrix in~(\ref{eq:matr}) become 
$v_{11}=v_{22}=1$, $v_{12}=v_{21}=0$. Hence, one can no longer obtain any 
information on $\tilde{\Delta}$  from~(\ref{eq:matr}), but natural 
conclusion that $b_{+}=a_{+}$ and $b_{-}=a_{-}$ follows due to that 
both branch cuts are on the same side from the contour of integration. 
Consequently, as already pointed out in the beginning of this section, such 
points $\omega_c$ are not remarkable.

\section{Possible astrophysical implications}

We believe that strongly magnetized inner parts of the jets can be 
directly connected to the central supermassive black hole by magnetic 
field lines. Let us estimate wether such jet can be force--free.   We 
assume the black hole have typical values of its mass and rotating 
parameter:  $M\simeq 10^8M_\odot$, $a=J/M\leq 1$.
On the distances from black hole comparable to the Schwarzschild radius
strong magnetic field is expected to be of the order of $10^4G$ 
(Begelman, Blandford \& Rees, 1984).
Than the electron 
density of $n_c\simeq 1cm^{-3}$ is enough to screen the longitudinal 
(along the magnetic field) electric field component so that the MHD 
approximation becomes possible.  In the inner part of the flow connected 
with the black hole by magnetic field lines the particle density $n$ 
cannot much exceed the value $n_c$  because the particles 
constrained to move along the magnetic surfaces do not escape the black 
hole and the only source for replenishing them is $e^+e^-$ pairs 
production. However, particles creation in the black hole magnetosphere 
seems to be possible only in the presence of the longitudinal 
electric field, which vanishes for $n\gg n_c$. Therefore, an equilibrium
between the particles outflow fed the jet and their creation is 
established for particle densities of the order of $n_c$. The necessity of 
particles creation in the magnetosphere of supermassive black hole was 
first pointed out by Blandford \& Znajek (1977). 
In this case the energy density of the 
magnetic and electric fields is $10^{13}$ times greater than the rest 
energy density of the $e^+e^-$ pairs. That makes the force-free 
approximation adequate in the vicinity of supermassive black hole.

Apparently, the force-free approximation is also 
valid for the inner parts of the jet, which are close to the axis of symmetry 
and are connected with the black hole.  
To show this consider the conservation of the particles flow and the 
conservation of the current in the process of possible recollimation of this
inner part of the jet from the sizes of the  order of the black hole event 
horizon to much greater.  Continuity of the particles flow implies that 
$\gamma n \propto 1/R^2$, where $R$ is the radius of the jet, $\gamma$ is 
the Lorentz factor of the plasma flow, $n$ is the particles concentration 
in the reference frame comoving with the plasma flow and the velocity of 
the particles is relativistic everywhere.  Conservation of the current 
flowing inside the jet leads to that the magnetic field after 
recollimation will be predominantly toroidal and scale as $B \propto 1/R$.  
We see therefore that the ratio $B^2/4\pi mn\gamma^2 \propto 1/\gamma$, so 
after recollimation to larger radii (say, parsecs) the jet remains to be 
force-free unless the Lorentz factor of the flow will not reach an 
unbelievably high value $10^{13}$. Therefore, we hope to apply our 
consideration for those jets, which are believed to be electron--positron, 
and are likely to be force--free. 

The remarkable feature of the dispersion curves $Re\, \omega(k)$ is that 
they have a minimum at some $k=k_{min}$ and $k_{min}\neq 0$ (see Figs.~5 
and~7).  At the same time waves damping $\mbox{Im}\, \omega(k_{min})$ is 
either small (it never exceeded 0.1 in our computations) or even equal to 
0 for modes with $m<0$.  Because of these, the perturbation with 
$k\approx k_{min}$ do not propagate since the group velocity $d\omega /dk$ 
vanishes for $k=k_{min}$.  In contrast to the waves having $k\neq k_{min}$ 
this wave packet undergoes only diffuse broadening due to the finite value 
of $d^2\omega /d k^2$ for $k=k_{min}$.  It means that such oscillations 
form the "standing wave" with the wave vector $k_{min}$. Such "standing 
wave" does not transmit any energy because of vanishing group velocity. 
We use quotation marks to name this phenomenon in order not to confuse it 
with the well known in many fields of physics standing waves, which are 
the linear superposition of two progressive waves. The amplitude of the 
"standing wave" will be larger than the amplitudes of other waves because 
it experiences a dispersion spreading only.  This phenomena is caused by 
the fact that the oscillations with wave vectors less and greater then 
$k_{min}$ propagate in the opposite directions. The phenomenon of 
"standing wave" takes place for axisymmetric perturbations as well~(IP).

After a long time after 
initial excitation  the pattern of disturbance is formed with the wave 
crests moving with the phase velocity $\omega(k_{min})/k_{min}$. In our 
numerical calculations this velocity was always greater than the velocity 
of light. A "standing wave" perturbation occupies progressively growing 
with time as $\propto t^{1/2}$ part of the jet. The edges of this pattern 
move with velocity $\propto t^{-1/2}$, which is less than the velocity of 
light. At the same time amplitude of the perturbation inside the "standing 
wave" region is decaying with time as $\propto t^{-1/2}$. It is impossible 
to transmit any information by moving wave crests faster than the 
speed of light.  If one has relativistic electrons 
emitting synchrotron radiation inside the magnetic configuration dealing 
with (which is the case for extragalactic jet), than this pattern will be 
visible.  This provide us with the new type of superluminal source. Now 
according to well known formula one can calculate the observable velocity 
of such superluminal source in the projection onto the plane of the sky 
\begin{equation}
V_{\rm obs}=\frac{v\sin \theta}{1-v/c\cos\theta},  \label{eq:obs}
\end{equation}
where $\theta$ is the angle between the jet axis and the 
line of sight of the observer, $v$ is the velocity of the superluminal 
source. In Fig.~10 the dependence of $V_{\rm obs}$ from $\theta$ is 
depicted. If $\theta <\theta^{\ast}={\rm arccos}c/v$ than $V_{\rm obs}<0$, 
i.e. the apparent motion of knots will be reversal, in the direction 
opposite to the wave vector ${\bf k}_{\mbox{min}}$. 
Observer will see superluminal motion ($\mid V_{\rm obs}\mid>c$) 
if $${\rm arccos}\frac{c}{\sqrt{2}v}-45^{\circ}<135^{\circ}-{\rm arcsin} 
\frac{c}{\sqrt{2}v}.$$  
Note, that the velocity, which enters into the effect of relativistic 
beaming, is the velocity of matter flow inside the jet rather than the 
velocity $v$. Consequently, the brightness of superluminally moving knots 
will not be affected by their motion.

The source of perturbations may be instability of accretion disk around 
the central black hole, which affects the magnetic field in the vicinity 
of a black hole, and, therefore, lead to the excitation of the disturbances 
at the base of the jet. If the jet is oriented close to the line of sight 
($\theta<\theta^{\ast}$), the observer will see chain of knots moving 
backward to the core. In the counterjet, if observable despite the fading 
due to the effect of relativistic beaming, knots will move outward from 
the core, as seen from Fig.~10. Another possibility is the excitation of 
perturbations in the jet at the hot spot, where the jet is ended. Though 
we consider only stationary equilibrium jet structure, the velocity of the 
advancing of the jet into the extragalactic medium is believed to be only 
mildly relativistic ($\sim 0.2\>c$) (see, e.g. Begelman, Blandford \& 
Rees, 1984), so one can admit that the phenomenon of "standing wave" may 
take place in that region as well. Here knots in the jet, provided that 
$\theta<\theta^{\ast}$, will move from the hot spot in the direction of 
the core, while in the counterjet they will move to the hot spot out of 
the core. If $\theta>\theta^{\ast}$ then the observable direction of 
motion for both jet and counterjet coincides with the projection of 
the real motion of knots onto the plane of sky, i.e. outward from the core 
and outward from the hot spot.

From expression~(\ref{eq:obs}) it follows that if 
$\theta\to\theta^{\ast}$, then $|V_{\rm obs}|\to\infty$ having different 
signs for $\theta<\theta^{\ast}$ and $\theta>\theta^{\ast}$. The equality 
of $V_{\rm obs}$ to infinity means that the whole jet splashes 
simultaneously throughout all it's length. In reality this can not come 
true. The jet radius changes along it's length, jet are usually slightly 
bent. These two reasons limit the value $|V_{\rm obs}|$ by some high but 
finite value. If the viewing angle of the jet $\theta$ is near 
$\theta^{\ast}$ and changes along the jet such that there exist pieces of 
the jet viewed at an angle less than $\theta^{\ast}$ and greater than 
$\theta^{\ast}$, then pairs of knots can be observed which either  
collide and vanish or emerge and move in the opposite directions, 
depending on whether $\theta$ changes along the direction of real motion 
of wave crests from being greater than $\theta^{\ast}$ to being less than 
$\theta^{\ast}$ (in the case of merging knots) or from being less than 
$\theta^{\ast}$ to greater than $\theta^{\ast}$ (in the case of newly 
born pairs of knots). When the wave crest of a "standing wave" pattern 
passes that piece of the slightly bent jet, where 
$\theta\approx\theta^{\ast}$, observer will see that the knot,
corresponding to that wave crest, is stretched along the jet and it's 
total luminosity (not the brightness) becomes considerably larger. Then 
the next wave crest passes through that place of the jet, and observer 
sees the next flash. This can be the reason for the quasi periodical 
splashing of the innermost parts of the jets with modest viewing angles 
$\theta\approx\theta^{\ast}$ of the order of $45^{\circ}$, while usually 
such bursts are explained by relativistic beaming of the moving knots 
along slightly curved trajectory in the jet having $\theta$ no more than a 
few degrees (see recent paper by Camenzind \& Krockenberger, 1992). 
However, careful examination of these effects needs considering physical 
equation describing disturbances propagation inside curved jet for finding 
dependence $v(\theta)$, not only kinematic picture. This is beyond the 
scope of the present paper.

It is the task for radioastronomers to 
detect such "natural" (not due to the effect of projection) superluminal 
motions. B\aa\aa th~(1992) reported about three epoch observation of one
component in 3C345 moving inward to the core, but he writes that the
significance of this observation still remains to be verified. Hardee
(1990) proposed another scenario which can lead to observation of backward
motions of the intersection points of the shocks in nonmagnetized jet.
In the frame of our model {\it periodical} structures moving backward to the
core may be observable while Hardee's model predicts an isolated knots.

\section{Summary}

A considerable amount of extragalactic jets are extremely well collimated 
and extends over the distances tens times longer than their radii. Bright 
well known example of such jet is one emerging from the galaxy M87. The 
problem of extraordinary stability is long standing problem. 
In this work we have shown numerically that a jet with a longitudinal 
current is stable within the force--free approximation for all velocities 
of longitudinal motion and for wide range of the velocities of rotation. 
We consider initial value problem for linearized set of relativistic 
equations describing disturbances propagation inside cylindrical jet. By 
using Laplace transformation we find asymptotical behaviour of 
perturbations over long time since initial excitation. The stability 
problem is reduced to eigenvalue problem for radial modes. It was shown 
that there exist Alfv\'en continuous spectrum of eigenfrequencies and 
discrete spectrum having the accumulation point where two Alfv\'en 
resonances coincide. Numerical calculation shows that all eigenfrequencies 
have been computed have negative or equal to zero imaginary part. This 
means the stability of the jet with respect to helicoidal perturbations as 
well as to axially symmetrical or pinch ones. The physical reason for 
stability is that there is a shear of the magnetic field because of 
changing the curling of the magnetic field lines with the radius (the 
absence of the shear would imply $\Omega^F(r)\propto 1/r$, which leads 
either to the value of rotational velocity $\Omega^Fr+K{B_0}_\phi$ being 
undefined on the symmetry axis of the jet, if $1+K{B_0}_z\neq 0$, or flow 
velocity of the matter being equal to the speed of light on the 
symmetry axis, if $1+K{B_0}_z=0$, both cases have no physical sense). 
Because of the fluid pressure is low compared to that of the 
electromagnetic field, even a small shear stabilizes the motion 
perpendicular to the magnetic field lines and prevent the development of 
instability.

We also find that the dispersion curves $\omega=\omega_{nm}(k)$ have 
minima for certain values of $k=k_{\rm min}$. This means that such 
oscillations form a "standing wave" with a wave vector $k_{\rm min}$. The 
wave crests of this "standing wave" are spirals moving along the jet with 
the velocity exceeding the speed of light. The amplitude of the "standing 
wave" will be larger than the amplitudes of other waves because it 
experiences a dispersion spreading only. An example, illustrating this 
behaviour of perturbation, was presented in IP. Relativistic particles 
emitting synchrotron radiation in the jet make the perturbed magnetic 
field visible for the observer. Thus we obtain a new type of superluminal 
sources, having their physical rather than only observable velocity
superluminal. The observation of the chains of knots moving backward to 
the core of the extragalactic radio source will be the evidence that the 
phenomenon described does take place in Nature. The phenomenon of 
"standing wave" may also account for the quasi periodical flashes of the 
central compact sources.

\begin{center}
{\Large \bf Acknowledgments}
\end{center}

One of us (VIP) thanks to I.A.~Subaev for providing him with the code of
Dormand--Prince 5-order method of numerical integration. This work was 
done under the partial support of Russian Ministry of Science on the 
programme "Astronomy" and of International Science Foundation.

\begin{center}
{\Large \bf Appendix A}
\end{center}
\setcounter{equation}{0}

\renewcommand{\theequation}{A\arabic{equation}}

Here we list the expression for different components of the perturbations 
of the magnetic field, electric field and velocity field in terms of 
${B_r}_1$ and $\xi_r$. 
Force-free approximation and ${B_z}_0=\hbox{const}$ are adopted.
We use the abbreviations~(\ref{26})--(\ref{28}) of the main text.
For magnetic field
\begin{eqnarray}
\lefteqn{{B_\phi}_1=\frac{i}{S}\Biggl[
\frac{d{B_r}_1}{dr}\left(
\frac{1}{F}\omega(\omega+k)\Omega^Fr-\frac{m}{r}\right)+{B_r}_1
\frac{d\Omega^F}{dr}\frac{r}{F^2}k\times\Biggr.} \nonumber\\
 & & \Biggl.(\omega+k)(\omega-k-m
\Omega^F)-\frac{m}{r^2}{B_r}_1+
{B_r}_1\frac{\Omega^F}{F}(\omega^2-
\omega k-2k^2)\Biggr],\label{A1}  
\end{eqnarray}
\begin{eqnarray}
{B_z}_1 & = & \frac{i}{S}\frac{\omega+k}{F}
\Biggl[\frac{d{B_r}_1}{dr}(\omega-k-m\Omega^F)-{B_r}_1\frac{d\Omega^F}{dr}
m\frac{\omega-k-m\Omega^F}{F}\Biggr. \nonumber\\
 & & \Biggl.+\frac{1}{r}{B_r}_1(\omega-k+m
\Omega^F)\Biggr].\label{A2}  
\end{eqnarray}
For electric field
\begin{eqnarray}
{E_r}_1 & = & \frac{i}{S}\Biggl[\frac{d{B_r}_1}{dr}
\frac{1}{F}\left(-\omega\frac{m}{r}+\Omega^Frk(\omega+k)+
\Omega^F\frac{m^2}{r}\right)+{B_r}_1\frac{d\Omega^F}{dr}\frac{r(\omega+
k)}{F^2}\times\Biggr. \nonumber\\
 & & \Biggl.(-k\omega+k^2-m\Omega^F\omega+m^2/r^2)+
{B_r}_1\frac{1}{F}
\left(-\frac{m\omega}{r^2}+\Omega^F\left(-2\omega^2-\omega k+k^2+
\frac{m^2}{r^2}\right)\right)\Biggr],\label{A3}  
\end{eqnarray}
\begin{equation}
{E_\phi}_1=-{B_r}_1\frac{\omega-m\Omega^F}{k+m\Omega^F},\label{A4}
\end{equation}
\begin{equation}
{E_z}_1={B_r}_1\Omega^Fr\frac{\omega+k}{k+m\Omega^F}.\label{A5}
\end{equation}
In the approximation of ideal force-free plasma the component of the 
velocity parallel to the magnetic field line does not enter into the basic 
equations~(\ref{5}),~(\ref{6}) of the main text and can be free. Therefore 
the component of the velocity first order perturbation parallel to zero 
order magnetic field remains also to be free. Another two components are 
\begin{equation}
{v_r}_1=-{B_r}_1\left(\frac{\omega-m\Omega^F}{k+m\Omega^F}-K{B_z}_0\right)
\frac{1}{{B_z}_0},\label{A6}
\end{equation}
\begin{eqnarray}
\lefteqn{{B_z}_0{v_\phi}_1-{B_\phi}_0{v_z}_1=\frac{i}{S}\Biggl[\left\{
\frac{d{B_r}_1}{dr}\left(\Omega^Fr(\omega+k)-\frac{m}{r}\right)-
{B_r}_1\left(\frac{m}{r^2}+\Omega^F(\omega+k)\right)\right\}\times\Biggr.}
\label{A7} 
\nonumber\\
& & \qquad\qquad\qquad\qquad
 \left( \frac{1}{F}(m\Omega^F-\omega)+K{B_z}_0\right)+  \\
& & \Biggl.{B_r}_1\frac{d\Omega^F}{dr}\frac{r}{F}(\omega+k) 
\left\{\omega-k+\frac{1}{F}\left(\omega m \Omega^F-
m^2{\Omega^F}^2-\frac{m^2}{r^2}\right)+K{B_z}_0(\omega-k-m
\Omega^F)\right\}\Biggr]. \nonumber
\end{eqnarray}

To obtain expressions for all components of the electric and magnetic 
fields and velocity  we substitute in (\ref{A1})--(\ref{A7}) for ${B_r}_1$ 
it's value from equation~(\ref{23}) of the main text, which express 
${B_r}_1$ via $\xi_r$. Thus, we lead to the following 
\begin{equation} 
{B_r}_1=i{B_z}_0F\xi_r,\label{A8} 
\end{equation}
\begin{equation}
{B_\phi}_1=-{B_z}_0\left\{\Omega^Fr\frac{d\xi_r}{dr}+\xi_r\frac{d}{dr}
(\Omega^Fr)+\frac{k}{S}\left[r\frac{d\xi_r}{dr}\left(\Omega^F(\omega+k)-
\frac{m}{r^2}\right)-\xi_r\left(\Omega^F(\omega+k)+\frac{m}{r^2}\right)
\right]\right\},\label{A9}
\end{equation}
\begin{equation}
{B_z}_1={B_z}_0\left\{-\left(\frac{d\xi_r}{dr}+\frac{1}{r}\xi_r\right)+
\frac{m}{rS}\left[r\frac{d\xi_r}{dr}\left(\Omega^F(\omega+k)-
\frac{m}{r^2}\right)-\xi_r\left(\Omega^F(\omega+k)+\frac{m}{r^2}\right)
\right]\right\},\label{A10}
\end{equation}
\begin{equation}
{E_r}_1={B_z}_0\left\{\Omega^Fr\frac{d\xi_r}{dr}+\xi_r\frac{d}{dr}(
\Omega^Fr)-\frac{\omega}{S}\left[r\frac{d\xi_r}{dr}\left(\Omega^F(\omega+k)-
\frac{m}{r^2}\right)-\xi_r\left(\Omega^F(\omega+k)+\frac{m}{r^2}\right)
\right]\right\},\label{A11}
\end{equation}
\begin{equation}
{E_\phi}_1=-i{B_z}_0(\omega-m\Omega^F)\xi_r,\label{A12}
\end{equation}
\begin{equation}
{E_z}_1=i{B_z}_0\Omega^Fr(\omega+k)\xi_r,\label{A13}
\end{equation}
\begin{equation}
{v_r}_1=-i{B_z}_0(\omega-m\Omega^F-K{B_z}_0F)\xi_r,\label{A14}
\end{equation}
\begin{eqnarray}
{B_z}_0{v_\phi}_1-{B_\phi}_0{v_z}_1=-{B_z}_0\Biggl\{r\xi_r\frac{d\Omega^F}
{dr}(1+K{B_z}_0)+\frac{1}{S}(m\Omega^F-\omega+K{B_z}_0F)\times
\Biggr. \nonumber\\ 
\Biggl.\left[r\frac{d\xi_r}{dr}\left(\Omega^F(\omega+k)- 
\frac{m}{r^2}\right)-\xi_r\left(\Omega^F(\omega+k)+\frac{m}{r^2}\right)
\right]\Biggr\}.\label{A15} 
\end{eqnarray}
The function $\xi_r(r)$ has logarithmic singularity at the point $r=r_A$,
which is the simple zero of $A$ (the case of a second order zero is 
considered in section~4), and is regular everywhere in the remaining part 
of complex plane $r$, except infinite point. Therefore, one can conclude 
from equations~(\ref{A8})--(\ref{A15}) the following behaviour for 
disturbances of different physical quantities near Alfv\'en resonant point 
$r=r_A$ ($x=(r-r_A)/r_A$) 
\begin{eqnarray}
 & & {B_r}_1\propto\log x, 
\qquad {B_\phi}_1\propto \frac{1}{x}, \qquad {B_z}_1\propto \frac{1}{x}; 
\nonumber\\ & & {E_r}_1\propto \frac{1}{x}, \qquad {E_\phi}_1\propto\log 
 x, \qquad {E_z}_1\propto\log x; \nonumber\\ & & {v_r}_1\propto\log x, 
 \qquad {B_z}_0{v_\phi}_1-{B_\phi}_0{v_z}_1 \propto \frac{1}{x}. \nonumber 
\end{eqnarray}
It might appear from equations~(\ref{A9})--(\ref{A11}), (\ref{A15}) 
that the quantities ${B_\phi}
_1$, ${B_z}_1$, ${E_r}_1$, ${B_z}_0{v_\phi}_1-{B_\phi}_0{v_z}_1$ are
singular at the $r=r_S$, which is the zero of $S$. However, this is not the 
case. Being regular at $r=r_S$ function $\xi_r$ can be expanded into 
power series in the neighbourhood of $r=r_S$.  
By expanding second order differential equation~(\ref{36}) on $\xi_r$ one
can find that the combination      
$rd\xi_r/dr(\Omega^F(\omega+k)-
m/r^2)-\xi_r(\Omega^F(\omega+k)+m/r^2)$, which enters into all 
expressions~(\ref{A9})--(\ref{A11}),~(\ref{A15}), becomes equal to $0$ 
when $S=0$.  Therefore, all physical quantities regular there. The only 
singularity may arise when $A=0$, i.e.  for modes from Alfv\'en continuum. 

\begin{center}
{\Large \bf Appendix B}
\end{center}
\setcounter{equation}{0}

\renewcommand{\theequation}{B\arabic{equation}}

In the general case, when ${B_z}_0\neq \mbox{const}$, one can perform 
calculations analogous to that was done in section~2 when deriving the 
radial eigenmode equations~(\ref{32}). The coefficient by the derivatives 
{$\displaystyle \frac{1}{r}\frac{d}{dr}(r\xi_r)$} and {$\displaystyle 
\frac{dp_{\ast}}{dr}$} is of special interest, because it contains the 
singular points of the system of two first order differential equations on 
the displacement and the disturbance of the total pressure. It occurred to 
be the following
\begin{equation}
{\cal A}={B_z}_0^2(\omega-m\Omega^F)^2+(\omega 
{B_{\phi}}_0+\Omega^Frk{B_z}_0)^2- 
\left(\frac{m}{r}{B_\phi}_0+k{B_z}_0\right)^2.\label{B1} 
\end{equation}
Zeroes of ${\cal A}(\omega_A,r)=0$ give us Alfv\'en resonant frequencies 
$\omega_A$ at a given value $r$. The expression~(\ref{B1}) is quadratic in 
$\omega$, therefore, for each $r$ the equation ${\cal A}=0$ has two 
solutions for $\omega_A(r)$. If we suppose ${B_z}_0=\mbox{const}$ and use 
for ${B_{\phi}}_0$ the expression~(\ref{11}) with the sign~'+', the  
equation~(\ref{B1}) will become
\begin{equation}
{\cal 
A}={B_z}_0^2(\omega+k)\left[\omega-k-2m\Omega^F+{\Omega^F}^2r^2(
\omega+k)\right]={B_z}_0^2(\omega+k)A.\label{B2}
\end{equation}
So ${\cal A}$ factorizes into two coefficient, one of which do not contain 
the $r$-dependence. These two coefficient are present in 
system~(\ref{32}).

Next, let us show that at any real $r$ the quadratic equation ${\cal A}=0$ 
has only real solutions $\omega_A$. The discriminator of this equation is
\begin{displaymath}
{\cal D}=(-m\Omega^F+\Omega^Frk\tau)^2-(1+\tau^2){\Omega^F}^2r^2\left(
k^2+\frac{m^2}{r^2}\right)+(1+\tau^2)\left(\frac{m}{r}\tau+k\right)^2,
\end{displaymath}
where $\tau={B_\phi}_0/{B_z}_0$. The solutions $\omega_A$ will be complex 
if and only if ${\cal D}<0$. The condition ${\cal D}<0$ transforms to
\begin{displaymath}
{\Omega^F}^2r^2\left(k+\tau\frac{m}{r}\right)^2>(1+\tau^2)
\left(k+\tau\frac{m}{r}\right)^2,
\end{displaymath}
which, in turn, means that ${E_r}_0^2>{B_0}^2$. In stationary ideal MHD 
configuration electric field can never exceed magnetic field. This would 
imply the velocity of the fluid ${\bf v}$ being greater that the velocity 
of light. Naturally, it can be easily shown that any solution of the 
equation~(\ref{10}), governing the stationary jet configuration, satisfy 
the requirements $|E|<|B|$ (see IP). Therefore, ${\cal D}$ must be 
nonnegative value and the solutions $\omega_A$ are always real. Thus, in 
the general case of force--free cylindrical equilibrium Alfv\'en continua 
are always real and do not lead to instability.    

\newpage

\begin{center}
{\Large References}
\end{center}
\parindent0pt

Abramowitz M., Stegun I.A., 1970, {\it Handbook of Mathematical 
Functions}, Dover, New York

Appert K., Gruber R., Vaclavik J., 1974, Phys. Fluids, 17, 1471

Appl S., Camenzind M., 1992, A\&A, 256, 354

B\aa\aa th L.B., 1992, Physica Scripta, T43, 57

Begelman M.C., Blandford R.D., Rees,M.J., 1984, Rev.Mod.Phys., 56, 255

Beskin V.S., Istomin Ya.N., Pariev V.I., 1992(a), in the book 
{\it Extragalactic Radio Sources---From Beams to Jets}, edited by 
J.~Roland, H.~Sol, G.~Pelletier. Cambridge University Press, Cambridge

Beskin V.S., Istomin Ya.N., Pariev V.I., 1992(b), AZh, 69, 1258 

Birkinshaw M., 1984, MNRAS, 208, 887

Blandford R.D., Pringle J.E., 1976, MNRAS, 176, 443

Blandford R.D., Znajek R.L., 1977, MNRAS, 176, 433

Bondeson A., Iacono R., Bhattacharjee A., 1987, Phys. Fluids, 30, 2167 (BIB)

Camenzind M., 1987, A\&A, 184, 341

Camenzind M., Krockenberger M., 1992, A\&A, 255, 59

Cohn H., 1983, ApJ, 269, 500

Contopoulos J., 1994, ApJ, 432, 508

Hain K., L\"ust R., 1958, Z. Naturforsch., 13a, 936

Hardee P.E., 1979, ApJ, 234, 47

Hardee P.E., 1987, ApJ, 313, 607

Hardee P.E., 1990, in the book {\it Parsec Scale Radio Jets}, edited by  
J.A.~Zensus, T.J.~Pearson, Cambridge University Press, Cambridge, p.~266

Istomin Ya.N., Pariev V.I., 1994, MNRAS, 267, 629 (IP)

Kadomtzev B.B., 1988, {\it Collective Phenomena in Plasma}, Nauka Press, 
Moscow 

Li Z., Chiueh T., Begelman M.C., 1992, ApJ, 394, 459

Lifshitz E.M., Pitaevskii L.P., 1979, Physical Kinetics, Nauka Press, 
Moscow

Lovelace R.V.E., Wang J.C., Sulkanen M.E., 1987, ApJ, 315, 504

Macdonald D.A., 1984, MNRAS, 211, 313

Mirabel I.F., Rodriguez L.F., 1994, Nature, 371, 46

Nitta S., Takahashi M., Tomimatsu A., 1991, Phys. Rev., 44, 2295

Novikov I.D., Frolov V.P., 1986, Physics of the Black Holes, 
Nauka Press, Moscow

Payne D.G., Cohn H., 1985, ApJ, 291, 655

Pelletier G., Pudritz R.E., 1992, ApJ, 394, 117

Timofeev A.V., 1970, Uspekhi Fiz. Nauk, 102, 185

Takahashi M., Nitta S., Tatematsu Ya., Tomimatsu A., 1990, ApJ, 363, 206

Thorne K.S., Price R.H., Macdonald D.A., 1986, Black Holes: 
The Membrane Paradigm, Yale University Press

Torricelli-Ciamponi G., Petrini P., 1990, ApJ, 361, 32

Turland B.D., Scheuer P.A.G., 1976, MNRAS, 176, 421

\newpage

\begin{center}
{\Large Figure captions}
\end{center}

{\bf Figure 1.} Schematic representation of the equilibrium stationary 
configuration of a jet with a uniform poloidal magnetic field $B_z$.
The frequency of rotation in the dimensionless units 
described in the beginning of Subection~2.1 is 
$\Omega^F=10(1-r^2)$.  The jet boundary for $r=1$ and three magnetic tubes 
for $r=1/4$, $2/3$ and $9/10$ are shown. The magnetic field lines are 
spiralling on a magnetic tube.  Since $\Omega^F(1)=0$, the total current 
through the jet is equal to zero and the magnetic field is purely poloidal 
both at the boundary and at the axis of symmetry. The curling of 
magnetic field lines is maximum for $r=1/\sqrt{3}$, decreasing for smaller 
and larger radii. The density of the poloidal current $j_z$ is negative 
when $r<1/\sqrt{2}$ and positive when $1>r>1/\sqrt{2}$. The electric field 
${\bf E}$ induced by jet rotation is radial. The plasma velocity ${\bf v}$ 
along the magnetic tube consists of two components: rotation with 
angular velocity $\Omega^F(r)$, and motion along the magnetic field lines 
with a speed ${\bf v}_\parallel=K{\bf B}$. We see that the rotation 
velocity $r\Omega^F$ can exceed the speed of light $c$ ( in our case the 
maximum value of $r\Omega^F$ is $20/3\sqrt{3}c$ at $r=1/\sqrt{3}$ ); 
nevertheless the quantity $v$ is restricted by $c$ due to the existence of 
a predominantly toroidal magnetic field. The dispersion curves 
$\omega=\omega_{nm}(k)$ for perturbations of this equilibrium state are 
plotted in the Figs.~5,6,7.

{\bf Figure 2.} Schematic picture, illustrating analytical properties of 
$\xi_{r\omega}$ in the complex $\omega$-plane. See explanation in the 
text. Brunch cuts are shown by thing solid line, the contour of 
integration shifted into lower $\omega$ half--plane breaks up into the 
circles around the poles $\omega=\omega_{nm}(k)$ and the part going round 
the branch points and along the attached branch cuts. This contour is 
shown by bold solid line with arrows.

{\bf Figure 3.} Schematic picture of the complex $r$-plane. The 
trajectories of $r_A(\omega)$ when changing $\omega$ along the paths shown 
in Fig.~2, are plotted by dashed lines, deformed contours of integration 
in cases {\sl C} and {\sl B} are plotted by bold solid lines, branch cuts 
(one for each case {\sl A}, {\sl B}, {\sl C} or {\sl D}) are shown by thin 
solid lines.

{\bf Figure 4.} The thin curves with arrows show the trajectories of 6 
points $r_A(\omega)$ when changing $\mbox{Im}\,\omega$ from $+\infty$ to 
some large negative value. The endpoints are marked by crosses. In all 
figures $\Omega^F=10(1-r^2)$, $m=1$, $k=-3$. Figures~(a), (b), (c), (d) 
and~(e) correspond to the continuation paths~{\sl A}, {\sl B}, {\sl C}, 
{\sl D}, and~{\sl E} in Fig.~2 respectively. $\omega_A(0)=17$, 
$\omega_A(1)=-3$, two complex conjugated $\omega_c$ are 
$\omega_c=3.3117\pm0.1564i$, $r_{\ast}$ is posed at $0.5$. In figure~(a) 
$\mbox{Re}\,\omega=-5$, in~(b) $\mbox{Re}\,\omega=5$, in~(c) 
$\mbox{Re}\,\omega=3.305$ , in~(e) $\mbox{Re}\,\omega=3.315$, in~(d) 
$\mbox{Re}\,\omega=17.5$. On figure~(d) 4 lateral loops 
are enlarged, because their dimensions are less than $0.01$ in reality.

{\bf Figure 5.} The dependence of the real part of $\omega_{nm}(k)$. 
$m=2$, $\Omega^F=10(1-r^2)$. Three lowermost branches of the dispersion 
relations are shown. They are distinguished from each other by an 
appropriate number. Straight lines are $\omega=k$ and 
$\omega=-k$.

{\bf Figure 6.} The dependence of the imaginary part of $\omega_{nm}(k)$. 
$m=2$, $\Omega^F=10(1-r^2)$. Three curves correspond to those shown on 
Fig.~5 and are indicated by appropriate numbers.

{\bf Figure 7.} The dependence of 
the real part of $\omega_{nm}(k)$. $m=-1$, $\Omega^F=10(1-r^2)$. 
Three lowermost branches of the dispersion 
relations are shown. They are distinguished from each other by an 
appropriate number. Straight lines are $\omega=k$ and $\omega=-k$.

{\bf Figure 8.} Radial dependence of the second mode for~(a) $m=-1$, 
$k=-0.4$, $\Omega^F=10(1-r^2)$, $\omega_{1-1}=6.86$ and~(b) 
$m=2$, $k=1$, $\Omega^F=10(1-r^2)$,
$\omega_{12}=8.54-5\cdot10^{-3}i$. The solid 
curve shows $\mbox{Re}\,\xi_r$ and the dashed line $\mbox{Im}\,\xi_r$. The 
solution was started so that $\xi_r$ is real as $r\to 0$. Unit of $\xi_r$ 
is arbitrary. Fig.~(c) shows radial dependence of the radial component of 
energy flux, calculated for mode~(a).

{\bf Figure 9.} Complex $x$ and $X$ plane, shown for the case 
$\arg\Delta=-\pi/3$. In Fig.~(a) $m{\Omega^F}'/D<0$, in Fig.~(b)
$m{\Omega^F}'/D>0$. Contour of integration is plotted by bold solid line,
branch cuts are shown by thin solid lines.

{\bf Figure 10.} The dependence of the observable 
velocity of the standing wave 
pattern $V_{\rm obs}$ on the angle $\theta$ between the jet axis and the 
line of sight of the observer. Phase velocity of the perturbation $v$ is 
chosen equal to $3c$. If the jet is pointed directly {\it to} the observer 
than $\theta=0$, if it is pointed directly {\it from} the observer than 
$\theta=180^{\circ}$. 

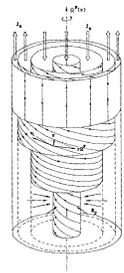
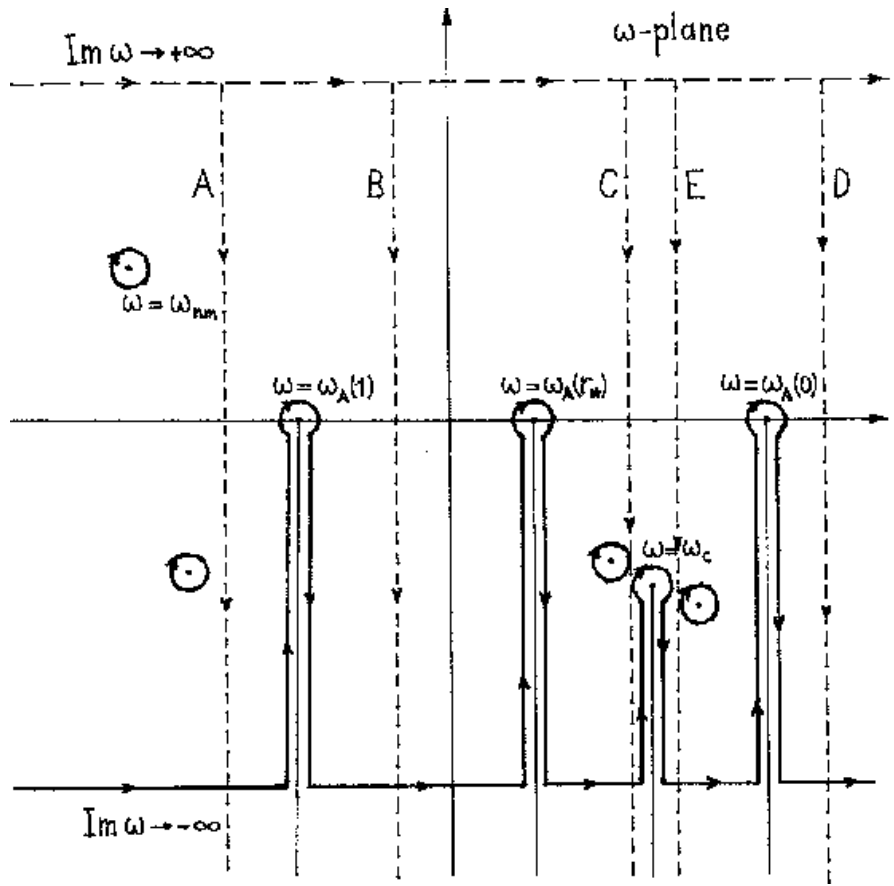
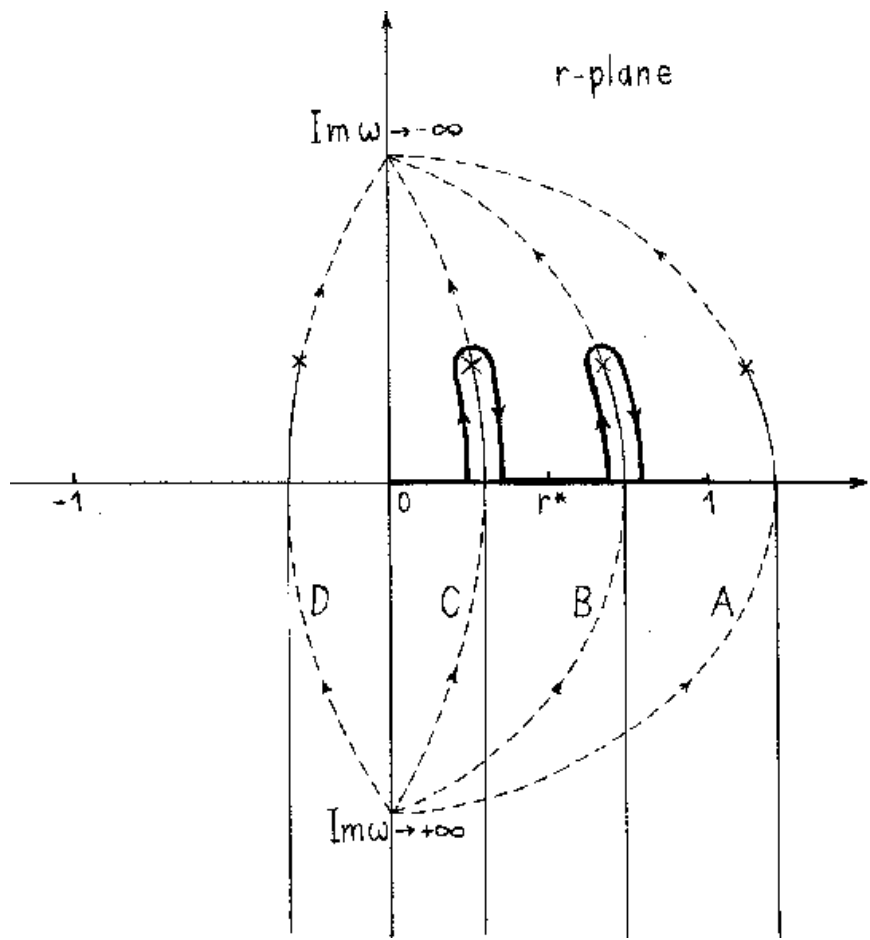
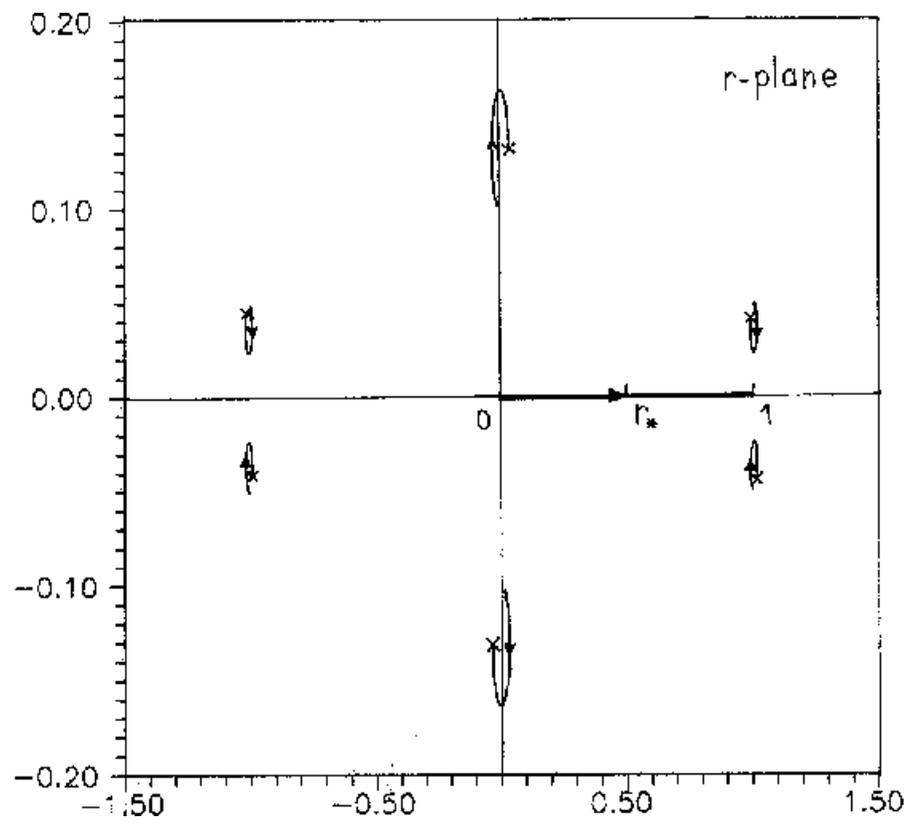
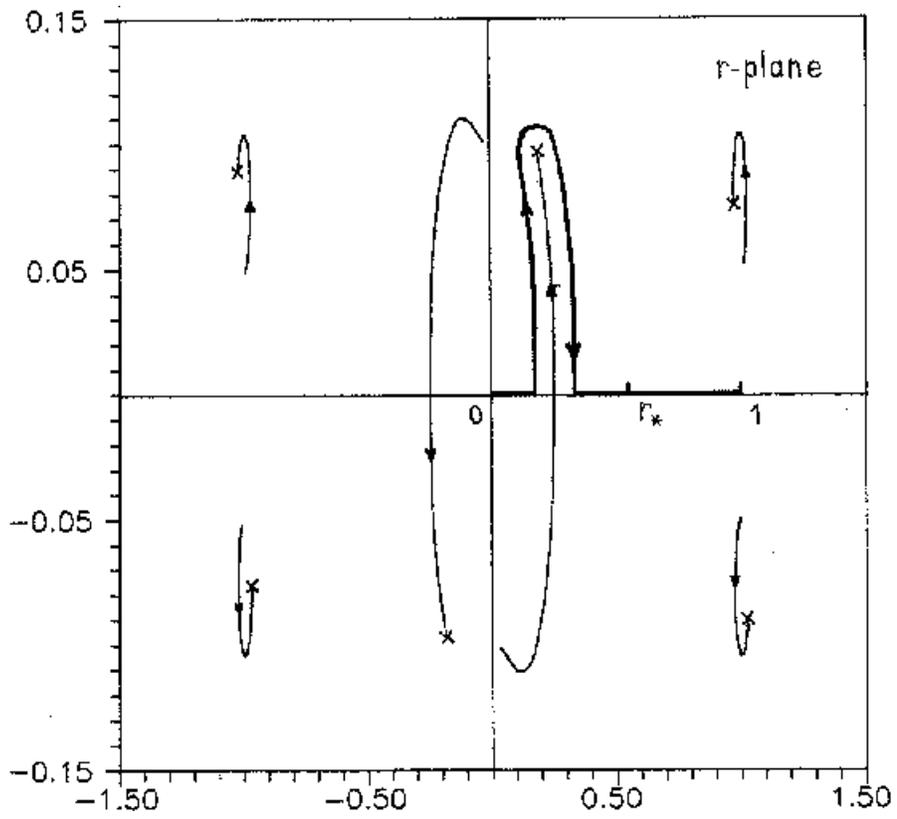
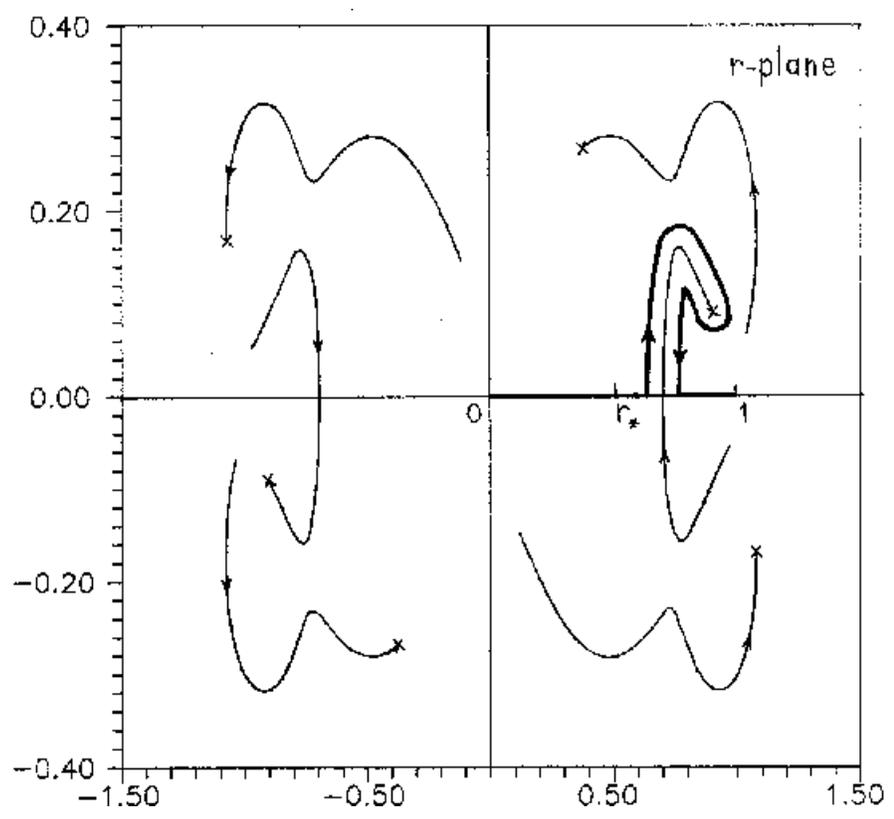
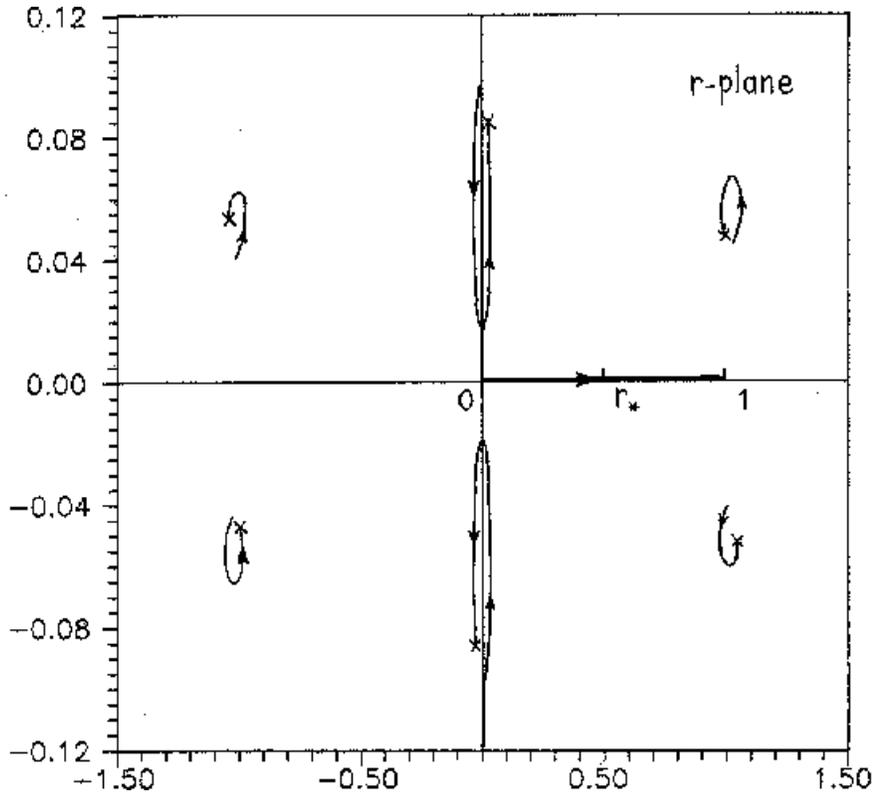
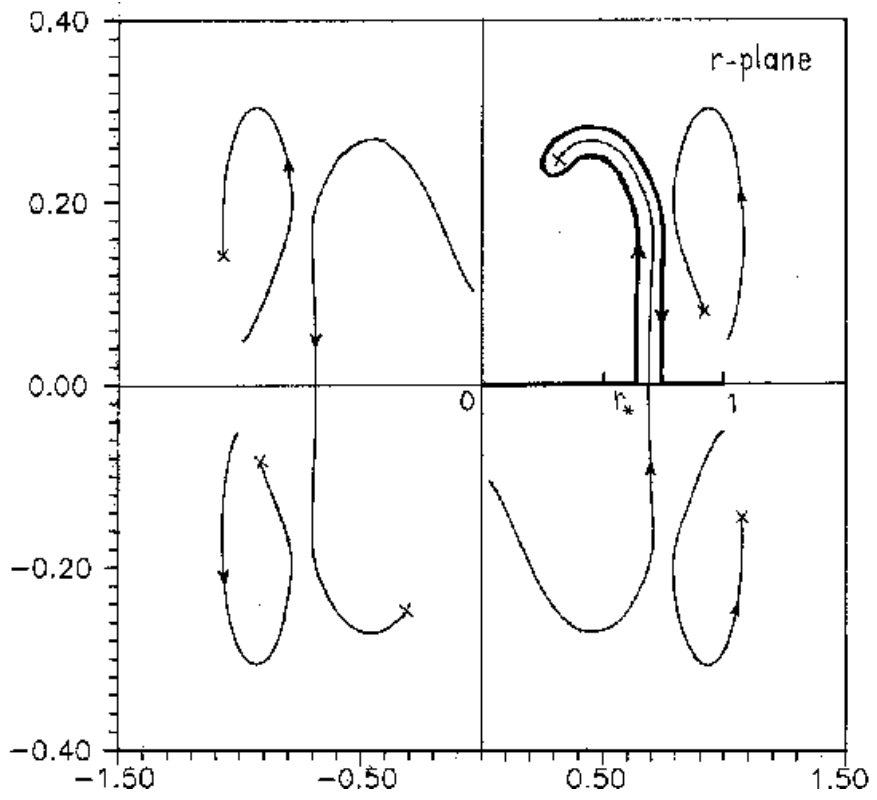
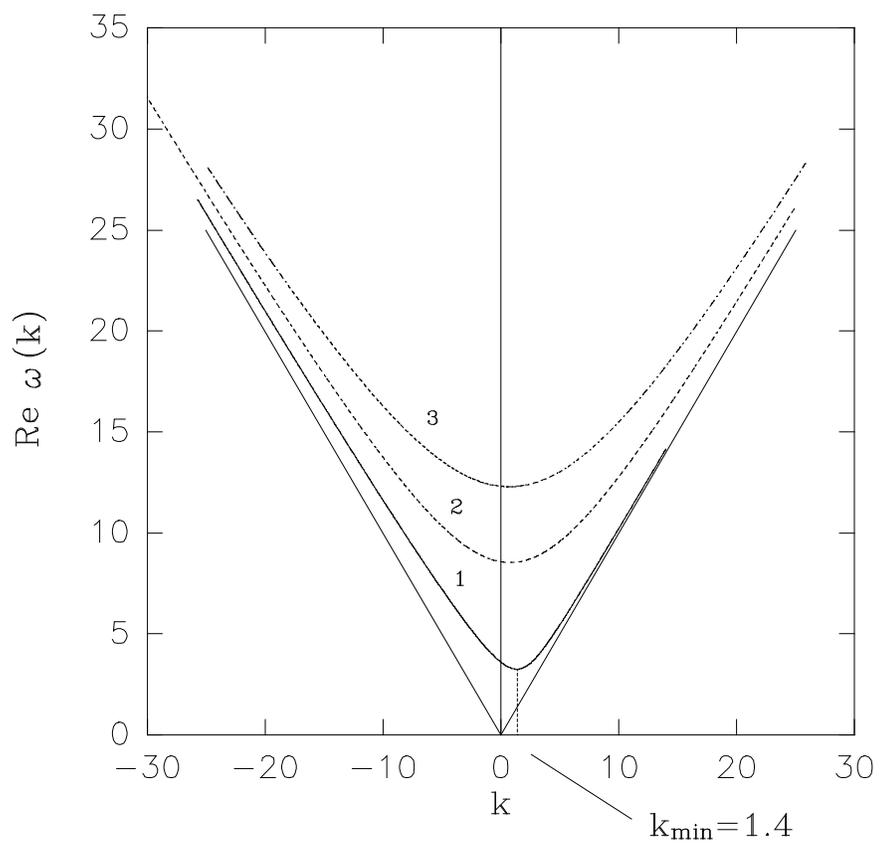
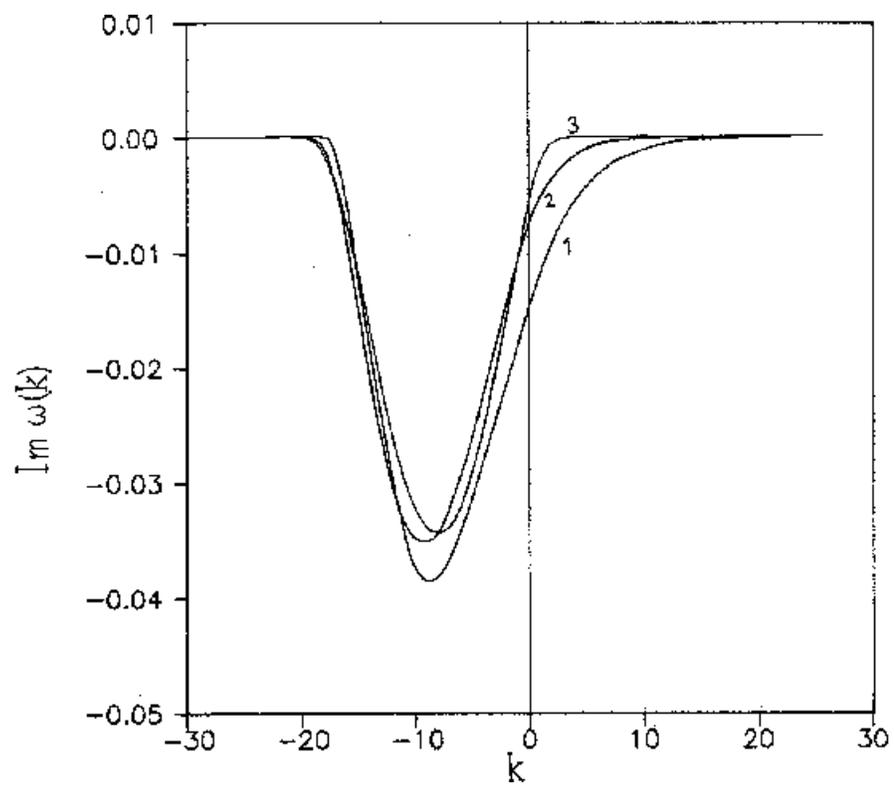
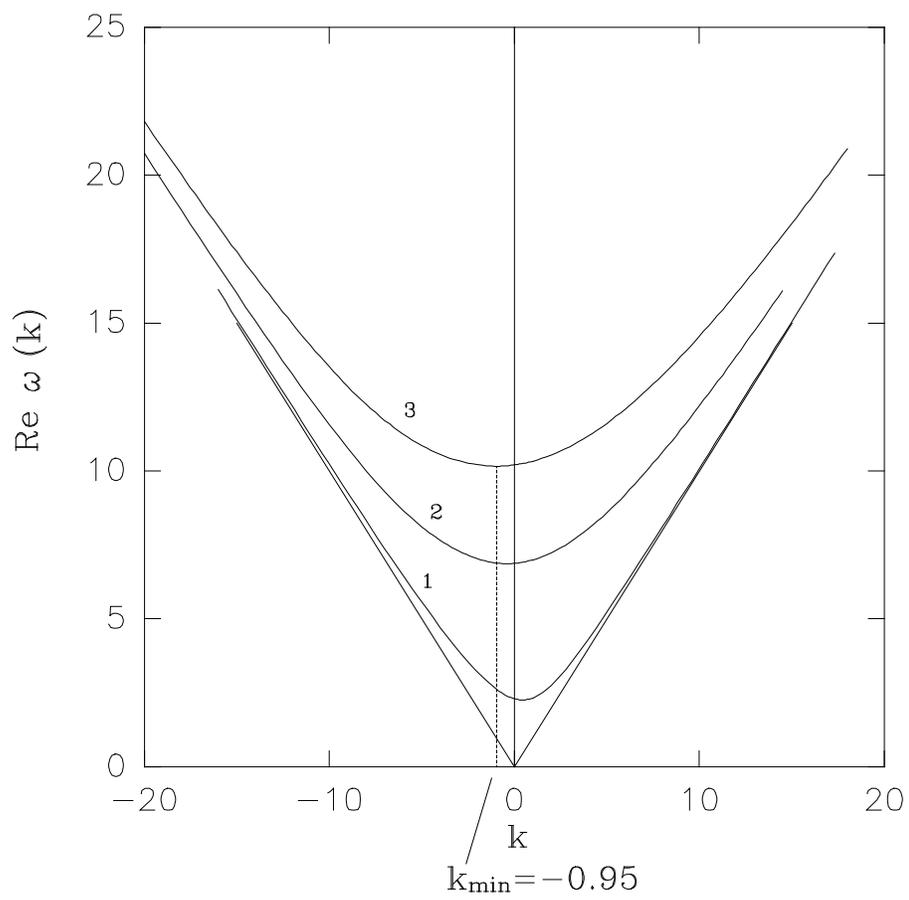
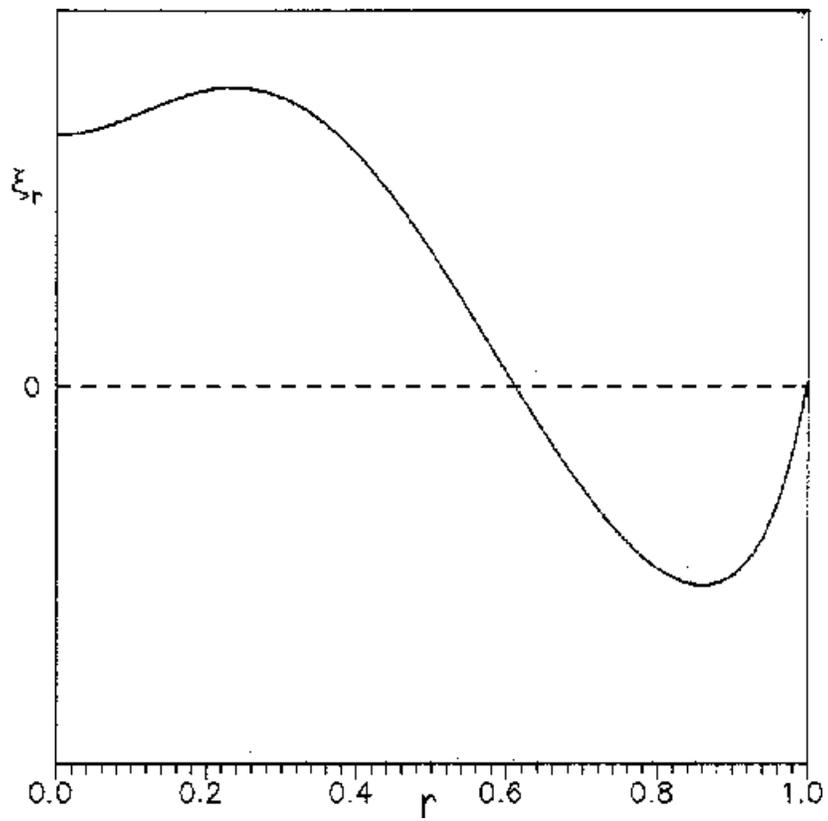
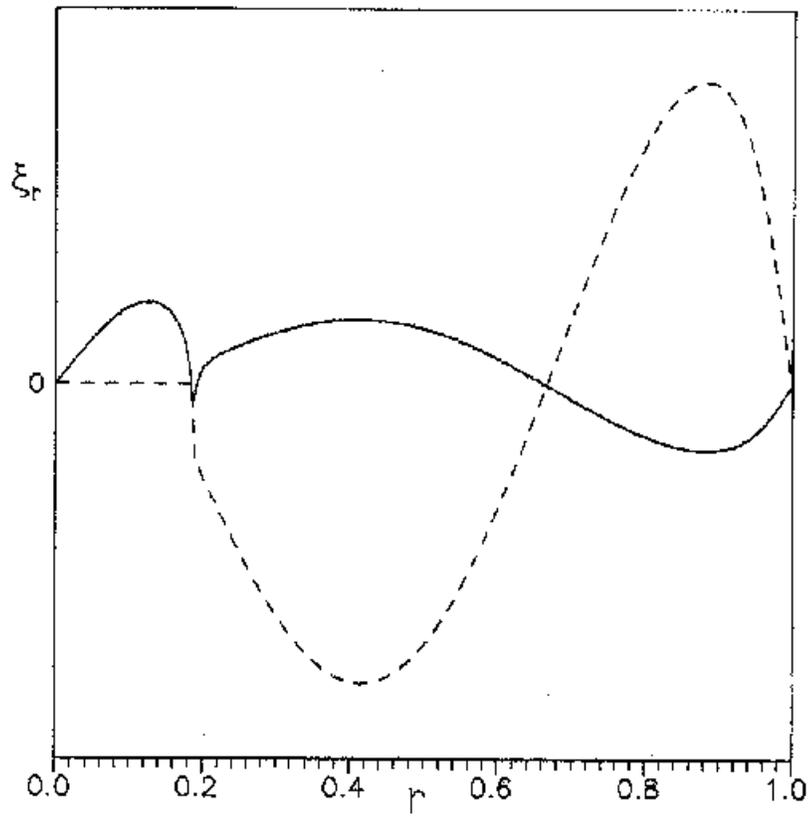
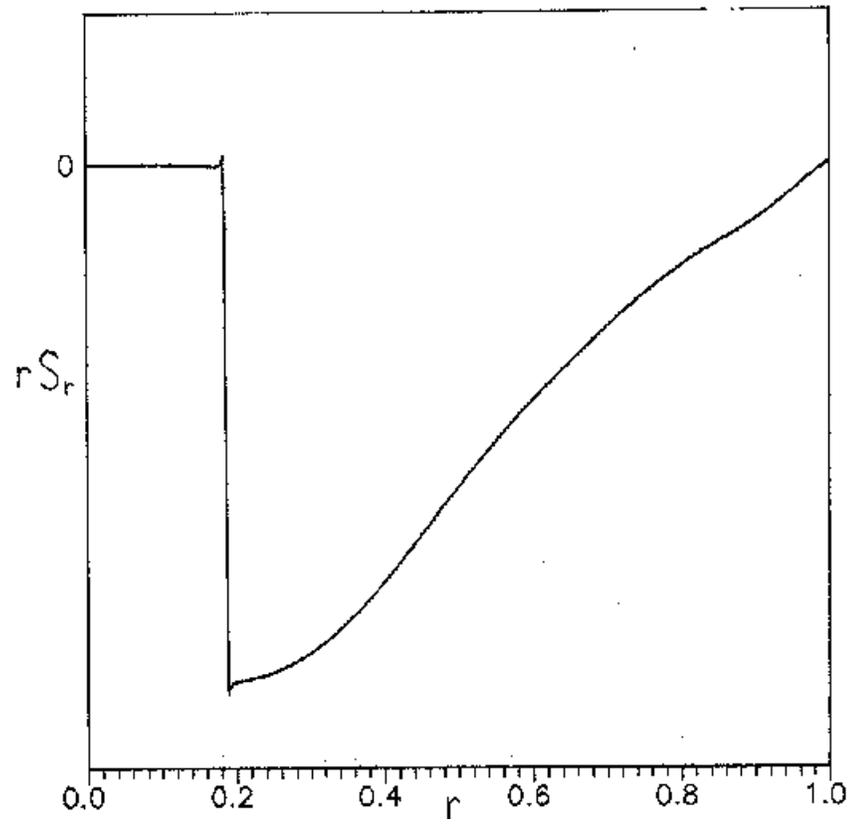
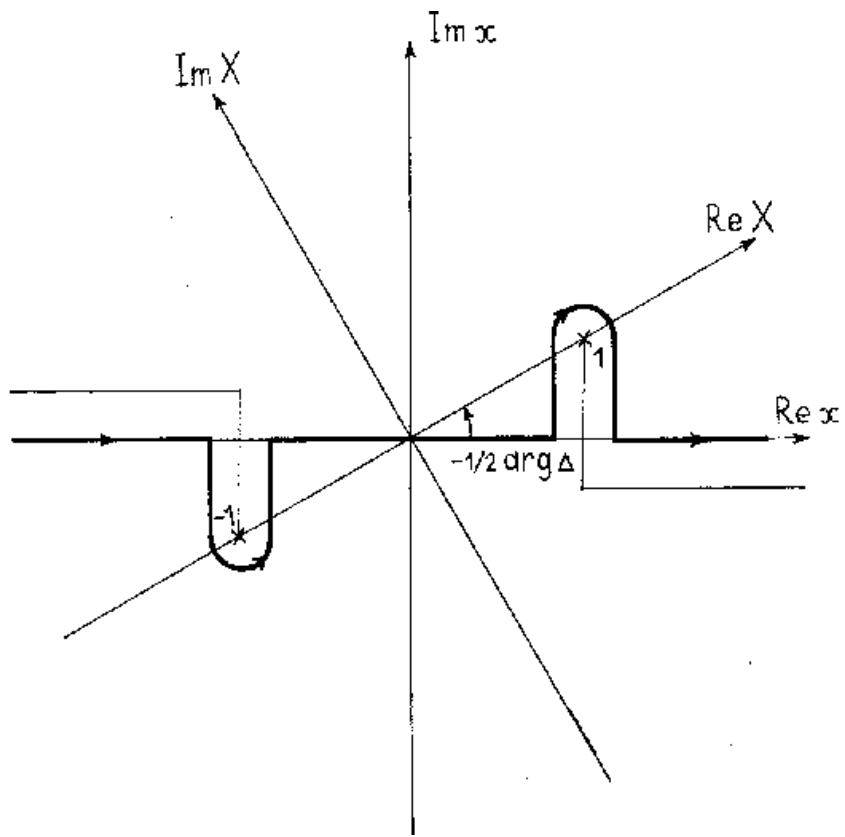
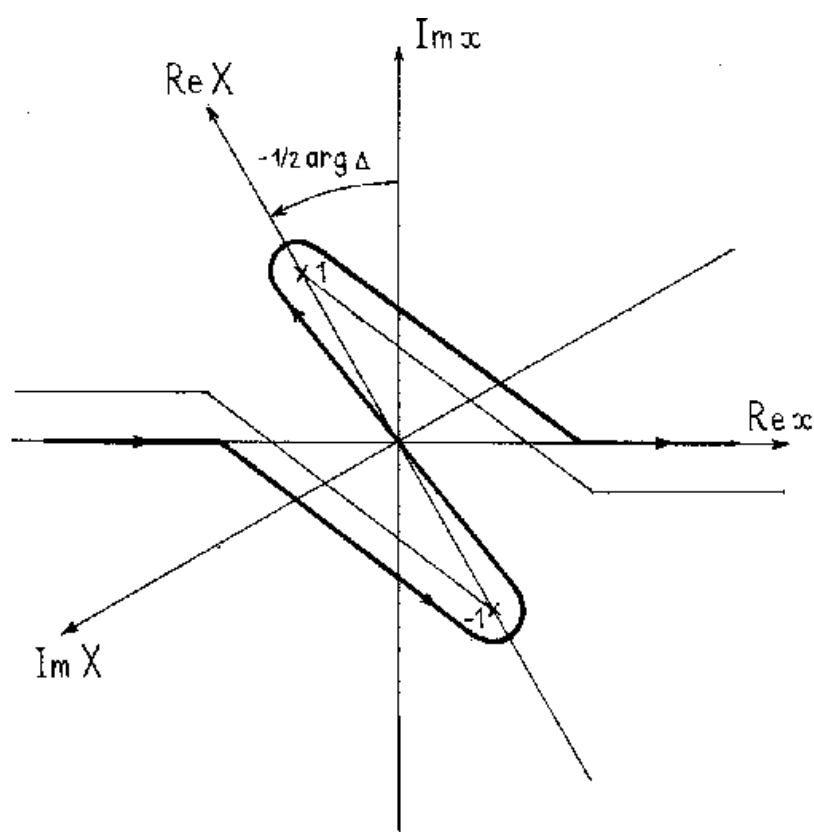
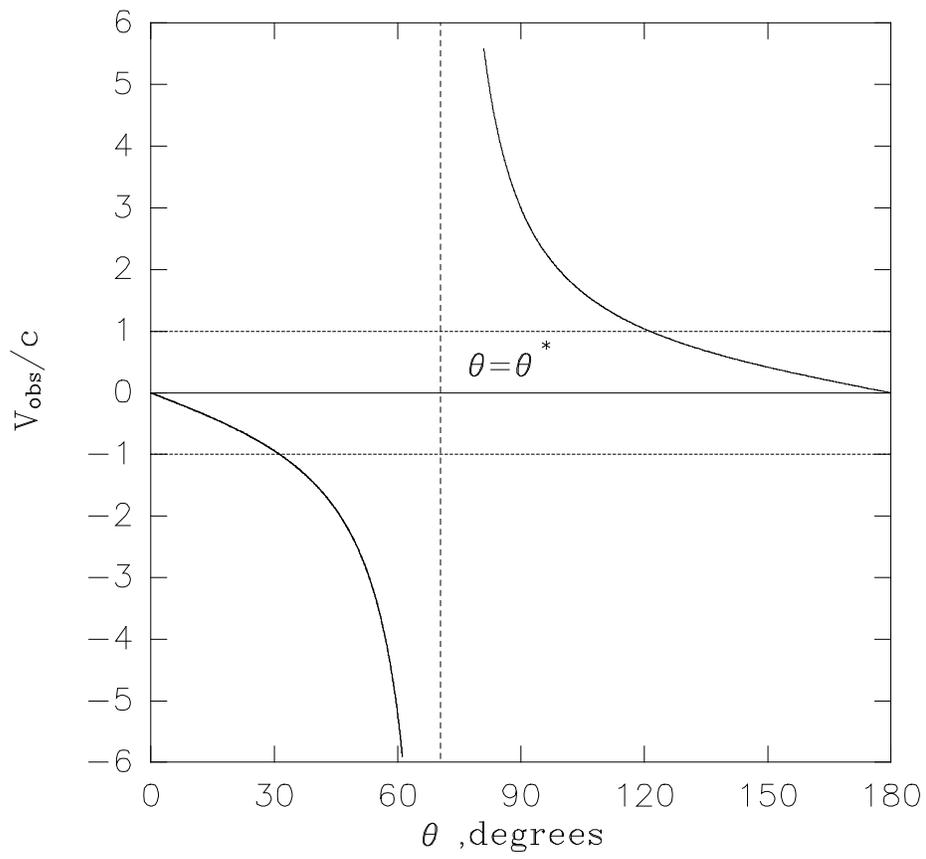
\end{document}